\documentclass[aps,groupedaddress,showpacs,showkeys]{revtex4}
\usepackage{amsmath,amscd,amssymb,epsfig}
\begin{document}

\bibliographystyle{apsrev}

\title{Theory of non-Markovian Stochastic Resonance}

\author{Igor Goychuk}
\email[]{goychuk@physik.uni-augsburg.de }

\altaffiliation{on leave from Bogolyubov
Institute for Theoretical Physics, Kiev, Ukraine}
\author{Peter H\"anggi}
\affiliation{Institute of Physics, University of Augsburg,
Universit\"atsstr. 1, D-86135 Augsburg, Germany}

\date{\today}

\begin{abstract}
We consider a  two-state model of non-Markovian
stochastic resonance (SR) within the framework of the
theory of renewal processes.
Residence time intervals are assumed to be mutually
independent
and characterized by some arbitrary
{\it non-exponential} residence time distributions which are
modulated in time by an externally applied signal.
Making use of a stochastic path integral approach we obtain general
integral equations governing the evolution of conditional
probabilities
in the presence of an input signal. These novel equations generalize
earlier integral renewal equations by Cox and others to the case of
driving-induced non-stationarity. On the basis of these new equations
a response theory
of two state renewal processes is formulated beyond the
linear response approximation.
Moreover, a general expression for the linear response function
is derived.
The connection of the developed approach with the
phenomenological theory of linear response for manifest non-Markovian
SR put forward in [ I. Goychuk and P. H\"anggi, Phys. Rev. Lett. {\bf 91}, 070601 (2003)]
is clarified
and its range of validity is scrutinized. The novel theory is then applied
to SR in symmetric non-Markovian
systems and to the class of single
ion channels possessing a fractal kinetics.

\end{abstract}



\maketitle

\section{Introduction}
The concept of Stochastic Resonance (SR) has been originally put forward
in order to explain the periodicity of glacial recurrences
on the Earth \cite{first}. It has gained, however,  an immense popularity in
the context of signal transduction in nonlinear stochastic
systems in physics and biology \cite{review1,review2,review3}.
Paradoxically enough, the detection
of beneficial input signals in the background stochastic fluctuations
of a signal-transmitting physical system can be improved upon
corrupting the information-carrying signal with input noise, or
upon raising the level of
intrinsic thermal noise. A first example of SR has been given
for a continuous state bistable dynamics
agitated by the thermal noise and  periodically
modulated by an external signal \cite{first}.
There exists a huge number of  systems in physics, chemistry, and
biology  which do
exhibit SR \cite{review1,review2,review3}. These
range from the
classical systems to the systems with distinct quantum
features \cite{GrifoniQSR}.

Experimentally, SR has been demonstrated in
various  macroscopic systems, see, e.g., in the  reviews \cite{review1,review3} and
the references therein. For a mesoscopic system containing a finite number of molecules
SR has been first demonstrated experimentally
in \cite{bezrukov}. The mesoscopic system in \cite{bezrukov}
consists of  dynamically self-assembled alamethicin
ion channels of variable size that are placed in a lipid
membrane. Up to this date, however, there remains the challenge to demonstrate
SR on the level of {\it single} stable molecules. Ion channels
of biological membranes  \cite{hille,neher} present
one of the  most appealing objects for such {\it single-molecular} studies.
The invention of patch clamp
technique by Neher and Sakmann in 1976 \cite{neher} made such
investigations possible. The single-molecular SR
experiments which have been performed under the conditions of
variable {\it intrinsic} thermal noise intensity  \cite{petrachi},
did not arrive at the convincing
conclusions. A recent {\it theoretical} study \cite{GH00} suggested a
parameter regime where SR effect should indeed occur for a Shaker
${\rm K}^{+}$ channel under physiological conditions when {\it external}
noise is added to the signal.
This issue has  further been
examined theoretically in \cite{ginzburg}. The present status calls 
for both theoretical, and experimental
further investigations. Particularly, the
presence of distinct memory effects in the dynamics of such single molecules
as ion channels \cite{west}
constitutes a major theoretical challenge.
The non-Markovian features caused by these memory effects
may be crucial for the occurrence  of Stochastic Resonance
on the level of single molecules.

The gross features of the observed bistable dynamics
can be captured by a two-state stochastic process
$x(t)$ that switches forth and back between two values $x_{1}$ and $x_{2}$
at random time points $\{ t_i\}$. Such a two-state random process
can be directly extracted from filtered experimental data and then
statistically
analyzed. Basically, the process $x(t)$ is characterized as follows:
The  sojourn in the state $x_1$ alternates randomly at $t_i$
into the sojourn in the state $x_2$, then $x(t)$ switches back to $x_1$ at
time $t_{i+1}$, and so on.
If the sojourn time intervals
$\tau_i=t_{i+1}-t_i$ are {\it independently} distributed (a condition
which we shall assume throughout the following),
such two-state
renewal processes are fully specified by two residence time distributions
(RTDs) $\psi_{1,2}(\tau)$ \cite{cox}. In the simplest case, which
corresponds
to the dichotomic  Markovian process (DMP),  both RTDs are strictly
exponential, i.e.,
$\psi_{1,2}(\tau)=\nu_{1,2}\exp(-\nu_{1,2}\tau)$,
where $\nu_{1,2}$ are the transition rates which equal the
inverse mean residence times (MRTs), which are given by
\begin{eqnarray}\label{MRT}
\langle \tau_{1,2}\rangle :=\int_{0}^{\infty}
\tau\psi_{1,2}(\tau)d\tau,
\end{eqnarray}
with $\nu_{1,2}=\langle \tau_{1,2}\rangle^{-1}$. The input
signal $f(t)$ causes the transition rates $\nu_{1,2}$ to be
time-dependent, i.e.,
$\nu_{1,2}\to\nu_{1,2}(t)$.
Moreover, the RTDs become functionals of the driving signal,
 \begin{eqnarray}\label{pdf1}
 \psi_{1,2}(t-t')\to
\psi_{1,2}(t,t')=\nu_{1,2}(t)\exp[-\int_{t'}^{t}\nu_{1,2}(\tau)d\tau].
\end{eqnarray}
As a consequence, the time-dependent probabilities
$p_{1,2}(t)$ of the states $x_{1,2}$
obey the master equations
\begin{eqnarray}\label{rate-eq}
\dot p_1(t)& = &-\nu_1(t)p_1(t)+\nu_2(t)p_2(t)\nonumber \\
\dot p_2(t)& = &\;\;\;\nu_1(t)p_1(t)-\nu_2(t)p_2(t)
\end{eqnarray}
with the signal-dependent rates which under
an adiabatic assumption  obey the rate law \cite{RMP},
\begin{equation}\label{rate-t}
\nu_{1,2}(t)=\nu_{1,2}^{(0)}\exp\Big (-[\Delta U_{1,2}\mp
\Delta x f(t)/2]/k_B T\Big ).
\end{equation}
In (\ref{rate-t}), $\nu_{1,2}^{(0)}$ are the frequency prefactors,
$\Delta U_{1,2}$ are the heights of the activation barriers,
$\Delta x:=x_2-x_1>0$ is the amplitude of fluctuations, $k_B$ is
the Boltzmann constant and $T$ is the temperature. For a weak periodic
signal
\begin{equation}\label{signal}
f(t)=f_0\cos(\Omega t),
\end{equation}
the use of Eqs. (\ref{rate-eq}),
(\ref{rate-t}) allows one to calculate within linear response theory the
{\it asymptotic}, long time response of the mean value
$\langle x(t)\rangle=x_1p_1(t)+x_2p_2(t)$
to $f(t)$, i.e.
\begin{equation}\label{response0}
\langle \delta x(t)\rangle=
f_0|\tilde\chi(\Omega)|\cos(\Omega t-\varphi(\Omega)),
\;\;{\rm as}\;\;t\to\infty\;.
\end{equation}
In (\ref{response0}), $\tilde\chi(\Omega)$ is the linear response
function in the frequency domain and $\varphi(\Omega)$ denotes
the phase shift.
The spectral amplification of signal,
$\eta=|\tilde\chi(\Omega)|^2$, exhibits the effect of
SR, i.e. a
bell-shaped dependence
versus increasing  intrinsic thermal noise strength which is measured by the temperature $T$
 \cite{review1}.

The above outlined two-state Markovian theory has been
put forward by McNamara and Wiesenfeld \cite{mcnamara}; this approach
has  proven very useful over the years as a basic, prominent
model for SR research \cite{review1}.
Remarkably enough,  this simple model allows one to unify
the various
kinds of SR -- periodic, aperiodic \cite{b3}, and even
non-stationary SR -- within a unifying framework of
information theory \cite{GH00}.

Many {\it observed} bistable stochastic processes
$x(t)$ are, however, truly not Markovian, as can be deduced from
the experimentally
observed RTDs. As a matter of fact, any deviation of RTDs from the
strictly exponential
form indicates a deviation from the
Markovian behavior \cite{remark0,boguna}.
The profoundly non-Markovian case emerges when at least one 
of the RTDs possesses a large (diverging) variance
${\rm var}(\tau_{1,2})=\int_{0}^{\infty}
\tau^2\psi_{1,2}(\tau)d\tau-\langle \tau_{1,2}\rangle^2\to\infty$.
The stochastic dynamics of single molecules is
especially interesting in this respect.
For example, the RTDs of the conductance fluctuations in
biological ion channels are in many cases not exponential
\cite{liebovitch,sansom,mercik}.
Usually, a sum of many exponentials, $\psi(\tau)=\sum_{i=1}^{N}c_i
\nu_i\exp(-\nu_i \tau),\;\sum_{i=1}^{N}c_i=1$
is needed to describe the experimental
data \cite{neher}.
Moreover, in some cases $\psi(\tau)$ can
well be
described by a stretched
exponential \cite{liebovitch},
or by a power law $\psi(\tau)
\propto 1/(b+\tau)^{\beta},\;\beta>0$ \cite{sansom,mercik}. The power law
is especially remarkable. For example, in Ref. \cite{mercik} such
a power law behavior has been found
for the closed time RTD of a large conductance (BK)
potassium channel with a power law exponent $\beta\approx 2.24$ yielding
formally ${\rm var}(\tau_{closed})=\infty$. This in turn implies that such
conductance fluctuations should exhibit a characteristic $1/f^{\alpha}$ noise
power spectrum $S(f)$
\cite{teich}. Indeed, this is the case of BK ion channel \cite{siwy},
as well as of some other ion channels \cite{bezrukov2000}.

What are the non-Markovian features of SR in similar systems?
We address this question below using the just described
non-Markovian generalization of McNamara-Wiesenfeld model
characterized by some
arbitrary non-exponential RTDs $\psi_{1,2}(\tau)$ and the
corresponding survival probabilities $\Phi_{1,2}(\tau)=
\int_{\tau}^{\infty}\psi_{1,2}
(\tau')d\tau'$ \cite{cox}. Similar models with alternating renewal
processes have been used previously in the SR theory for some
particular stochastic dynamics contracted
to the two-state dynamics
\cite{melnikov,lindner}. Moreover, the class of
colored noise driven Stochastic Resonance \cite{new1}
is also intrinsically non-Markovian. All these prior studies have been restricted, however,
to situations with finite memory effects on a finite time scale.
A truly non-Markovian situation emerges when the
memory effects extend practically to infinity, exhibiting a scale free,  weak
power law decay.
A phenomenological linear response
theory of such genuine non-Markovian SR
(which does not presume a knowledge
of the underlying microscopic dynamics) has been put
forward recently in Ref. \cite{goychuk03}.
The present work provides further details and, additionally,
presents a more general
framework for the non-Markovian SR theory which extends
beyond the linear response description.

\section{General theory}

\subsection{Two-state renewal process}

To start, let us consider a two-state renewal process (TSRP)
$x(t)$ which
takes initially, at time $t_0$, the value $x_1$, or the value $x_2$
with the probability
$p_1(t_0)$, or $p_2(t_0)$, correspondingly. At a random time point $t_1$
the process switches its current state with the probability one into
another state and stays there until some next random time
point $t_{2}$. Then, the renewal  process proceeds further
in time in the same manner.
The survival probability
to remain in the state $1$, or the state $2$ for the time
$\tau_i=t_{i+1}-t_i$ is
$\Phi_{1}(\tau_i)$, or $\Phi_{2}(\tau_i)$, correspondingly. These
two survival probabilities completely specify the considered TSRP
\cite{cox}. The functions $\Phi_{1,2}(\tau)$
must satisfy the following obvious restrictions:
(i) $0\leq\Phi_{1,2}(\tau)\leq 1$, (ii)
$\Phi_{1,2}(\tau+\Delta\tau)\leq \Phi_{1,2}(\tau), \;\Delta\tau>0$ (non-increasing
function of time) (iii) $\Phi_{1,2}(0)=1$;
(iv) $\lim_{\tau\to\infty} \Phi_{1,2}(\tau)=0$, but are
otherwise arbitrary. One example is given by  the
stretched exponential law, or Weibull distribution:
\begin{eqnarray}\label{Weibull}
\Phi(\tau)=\exp\Big [-[\Gamma(1+1/a)\nu \tau]^a
\Big],\;0<a<1.
\end{eqnarray}
In (\ref{Weibull}), $\nu=1/\langle \tau\rangle$ is a rate parameter
having the meaning of inverse mean residence time (MRT)
and $\Gamma(x)$ denotes the
gamma-function. Moreover, the power law dependence,
\begin{eqnarray}\label{Pareto}
\Phi(\tau)=\frac{1}{[1+\nu \tau/\gamma]^{1+\gamma}},\;\gamma>0.
\end{eqnarray}
corresponds to the Pareto
distribution. Both Weibull and Pareto distributions typify
the so-called fractal dependencies. In particular, such distributions
have been
detected for several different types of ion channels
\cite{liebovitch,mercik}. An interesting feature of the Pareto
distribution is that for $0<\gamma<1$ it displays a diverging
variance, $var(\tau)=\infty$, whereas the MRT
$\langle \tau\rangle$ is finite. The closed time-intervals of a large
conductance potassium ion
channel studied in Ref. \cite{mercik} seems to obey (\ref{Pareto})
with $\gamma\approx 0.24$.
Other fractal-like distributions can be constructed from the
expansion over exponentials
\begin{eqnarray}\label{gen}
\Phi(\tau)=\sum_{i=1}^{\infty}c_i\exp(-\nu_i \tau),\;\sum_i c_i=1,
\end{eqnarray}
assuming some recurrence scaling relations among the rate constants
$\{\nu_i \}$, e.g., $\nu_{i+1}=a\nu_i$,
and among the expansion coefficients $\{c_i \}$, e.g., $c_{i+1}=b c_i$,
with some structural constants $0<a<1,0<b<1$  \cite{palmer,west,hughes}.
If the hierarchy of rate constants is obtained from a fundamental rate
constant
$\nu_0$ applying a recurrence scaling relation similar to one given above,
the corresponding distribution
can be characterized as a fractal in time. If the whole hierarchy is
produced by a more complicated scaling law involving two, or more
independent fundamental rate constants,
the distribution is multi-fractal.
The corresponding stochastic processes can be referred to as  fractal
renewal processes \cite{teich}.
Such random processes presently attract renewed attention in physics and in
mathematical biology  \cite{west}.

The negative time-derivative
\begin{eqnarray}\label{survival1}
\psi_{1,2}(\tau)=-\frac{d \Phi_{1,2}(\tau)}{d\tau}
\end{eqnarray}
yields the corresponding residence time distributions \cite{cox}.
Next,
let us assume that a number of alternations occurred before the
starting time
point $t_0$
and the considered process became
homogeneous in time {\it before} the observation
started at $t_0$.
Then, for such persistent, {\it time-homogeneous}
process the RTDs of the {\it first}
time interval $\tau_0=t_1-t_0$,
$\psi^{(0)}_{1,2}(\tau)$ must differ from $\psi_{1,2}(\tau)$
\cite{cox,hughes,tunaley,sher,boguna}.
Namely \cite{remark1},
\begin{eqnarray}\label{pdf2}
\psi_{1,2}^{(0)}(\tau)=\frac{\Phi_{1,2}(\tau)}{\langle
\tau_{1,2}\rangle},
\end{eqnarray}
where $\langle \tau_{1,2}\rangle$ is given  by Eq. (\ref{MRT}).
The corresponding survival probability of the first residence
time-interval reads
\begin{eqnarray}\label{surv2}
\Phi_{1,2}^{(0)}(\tau)=\frac{\int_{\tau}^{\infty}
\Phi_{1,2}(t)dt}{\langle
\tau_{1,2}\rangle}.
\end{eqnarray}
Moreover, if to choose $p_{1,2}(t_0)$ as the stationary values,
$p_{1,2}(t_0)=p_{1,2}^{st}$, the
considered persistent process is {\it stationary}.
From Eq. (\ref{pdf2}) it follows that the two-state
renewal process (TSRP) can be stationary only
if the two mean residence times,
$\langle \tau_{1}\rangle$ and $\langle \tau_{2}\rangle$,
are finite. A diverging mean residence time leads to anomalously
slow diffusion (subdiffusion)
in the multi-state case \cite{hughes,sher,shlessinger}; such
situations will not be addressed with this work.

When a time-dependent input signal is switched on, the driven
TSRP becomes a non-stationary process and the corresponding  survival
probabilities depend not only on the length of time intervals,
but also on the initial time instant $t'$ of any considered residence time
interval, i.e. $\Phi_{1,2}(t-t')\to\Phi_{1,2}(t,t')$.
The  residence time
distributions are then accordingly given by
\begin{eqnarray}\label{survival}
\psi_{1,2}(t,t')=-\frac{d \Phi_{1,2}(t,t')}{dt}.
\end{eqnarray}
The corresponding conditional  survival probabilities can be defined
as $\Phi_{1,2}(\tau|t'):=\Phi_{1,2}(t'+\tau,t')$ (here the
condition is different from that used in footnote \cite{remark1} --
in the absence of signal --
notwithstanding the use of identical notations).
The particular choice, $\Phi_{1,2}(t,t')=\exp\Big(-\int_{t'}^{t}
\nu_{1,2}(\tau)d \tau\Big)$, leads to Eq. (\ref{pdf1}) -- the only
choice which is consistent with the Markovian
assumption \cite{remark0}.
In the non-stationary driven case,  the distinction
between $\Phi^{(0)}_{1,2}(t,t')$ and $\Phi_{1,2}(t,t')$,
$\psi^{(0)}_{1,2}(t,t')$ and $\psi_{1,2}(t,t')$
is not necessary. Nevertheless, we keep formally this distinction in the
following, because when the driving is being switched off, the process
$x(t)$ relaxes to its  stationary state. This
distinction becomes very important in order to construct the
evolution operator for time-homogeneous initial preparations.

\subsection{Integral equations of non-stationary renewal theory}

Our immediate goal is to obtain the evolution equations for
the considered stochastic process: we are looking for
the forward evolution operator $\mathbf \Pi(t|t_0)$ (or the matrix of
conditional probabilities) connecting
the probability vector $\vec p(t)=[p_1(t),p_2(t)]^T$ at two different
instants of time $t$ and $t_0$; i.e.,
\begin{eqnarray}\label{start}
\vec p(t)={\mathbf \Pi} (t|t_0)\vec p(t_0)\,.
\end{eqnarray}
This evolution operator can be explicitly constructed by considering
the contributions of all possible stochastic paths leading
from $\vec p(t_0)$ to  $\vec p(t)$. To start, let us separate these
contributions as follows
\begin{eqnarray}\label{structure}
{\mathbf \Pi} (t|t_0)=\sum_{n=0}^{\infty}{\mathbf \Pi} ^{(n)}(t|t_0),
\end{eqnarray}
where the index $n$ denotes the number of alternations that
occurred during
the stochastic evolution. The contribution with no alternations
obviously reads,
\begin{eqnarray}
{\mathbf \Pi} ^{(0)}(t|t_0)=\left [ \begin{array}{cc}
 \Phi_1^{(0)}(t,t_0) &  0 \\
 0 & \Phi_2^{(0)}(t,t_0)
\end{array} \right ]\;.
\end{eqnarray}
Stochastic paths with a single alternation  contribute
as
\begin{eqnarray}
{\mathbf \Pi} ^{(1)}(t|t_0)=\int_{t_0}^{t}dt_1{\mathbf P}(t,t_1){\mathbf
F}^{(0)}(t_1,t_0),
\end{eqnarray}
where
\begin{eqnarray}
{\mathbf P}(t,t_0)=\left [ \begin{array}{cc}
 \Phi_1(t,t_0) &  0 \\
 0 & \Phi_2(t,t_0)
\end{array} \right ]
\end{eqnarray}
and
\begin{eqnarray}
{\mathbf F}^{(0)}(t,t_0)=\left [ \begin{array}{cc}
 0 & \psi_2^{(0)}(t,t_0)  \\
  \psi_1^{(0)}(t,t_0) & 0
\end{array} \right ]\;.
\end{eqnarray}
 Next, the paths with two alternations
contribute to Eq. (\ref{structure}) as
\begin{eqnarray}
{\mathbf \Pi}
^{(2)}(t|t_0)=\int_{t_0}^{t}dt_2\int_{t_0}^{t_2}dt_1{\mathbf
P}(t,t_2){\mathbf F}(t_2,t_1){\mathbf F}^{(0)}(t_1,t_0),
\end{eqnarray}
where
\begin{eqnarray}\label{endstructure}
{\mathbf F}(t,t_0)=\left [ \begin{array}{cc}
 0 & \psi_2(t,t_0)  \\
  \psi_1(t,t_0) & 0
\end{array} \right ]\;.
\end{eqnarray}
Contributions with higher $n$ are constructed along the same
line of reasoning.

This representation of the evolution operator
${\mathbf \Pi}(t|t')$
in terms of an infinite sum over the stochastic paths is exact, although
not very useful in practice. The structure of the infinite series in Eqs.
(\ref{structure})--(\ref{endstructure}) implies, however, the following
representation
\begin{eqnarray}\label{inteq1}
{\mathbf \Pi} (t|t_0)={\mathbf \Pi} ^{(0)}(t|t_0)+
\int_{t_0}^{t}dt_1{\mathbf P}(t,t_1){\mathbf
G}(t_1,t_0),
\end{eqnarray}
where the unknown auxiliary matrix function $\mathbf G(t,t_0)$ satisfies the
matrix integral
equation
\begin{eqnarray}\label{inteq2}
{\mathbf G} (t,t_0)={\mathbf F} ^{(0)}(t,t_0)+
\int_{t_0}^{t}dt_1{\mathbf F}(t,t_1){\mathbf
G}(t_1,t_0).
\end{eqnarray}
The equivalence of Eqs. (\ref{structure})--(\ref{endstructure})
and Eqs. (\ref{inteq1})--(\ref{inteq2}) can be readily checked by solving
Eq.(\ref{inteq2}) with the method of successive iterations.

In components, Eq. (\ref{inteq1}) reads
\begin{subequations}
\begin{equation}\label{P11}
\Pi_{11}(t|t_0) =\Phi^{(0)}_1(t,t_0)+
\int_{t_0}^{t}\Phi_1(t,t_1)G_{11}(t_1,t_0)dt_1\;,
\end{equation}
\begin{equation}\label{P22}
\Pi_{22}(t|t_0) =\Phi^{(0)}_2(t,t_0)+
\int_{t_0}^{t}\Phi_2(t,t_1)G_{22}(t_1,t_0)dt_1\;,
\end{equation}
\begin{equation}\label{P12}
\Pi_{12}(t|t_0) =\int_{t_0}^{t}\Phi_1(t,t_1)G_{12}(t_1,t_0)dt_1\;,
\end{equation}
\begin{equation}\label{P21}
\Pi_{21}(t|t_0)
=\int_{t_0}^{t}\Phi_2(t,t_1)G_{21}(t_1,t_0)dt_1\;.
\end{equation}
\end{subequations}
It is worth to notice that the set of Eqs. (\ref{P11})-(\ref{P21})
is not independent. The conservation of probability implies that
\begin{eqnarray}\label{conserv}
\Pi_{11}(t|t_0)+\Pi_{21}(t|t_0)& = &1\;, \nonumber \\
\Pi_{22}(t|t_0)+\Pi_{12}(t|t_0) & = &1.
\end{eqnarray}
The consistency of Eqs. (\ref{P11})-(\ref{P21}) with the
conservation law, Eq. (\ref{conserv}),
can be checked readily.
The matrix integral equation(\ref{inteq2}) reads in components
\begin{subequations}
\begin{equation}\label{G11}
G_{11}(t,t_0) =\int_{t_0}^{t}\psi_2(t,t_1)G_{21}(t_1,t_0)dt_1\;,
\end{equation}
\begin{equation}\label{G22}
G_{22}(t,t_0) =\int_{t_0}^{t}\psi_1(t,t_1)G_{12}(t_1,t_0)dt_1\;,
\end{equation}
\begin{equation}\label{G12}
G_{12}(t,t_0) =\psi^{(0)}_2(t,t_0)+\int_{t_0}^{t}
\psi_2(t,t_1)G_{22}(t_1,t_0)dt_1\;,
\end{equation}
\begin{equation}\label{G21}
G_{21}(t,t_0)
=\psi^{(0)}_1(t,t_0)+\int_{t_0}^{t}\psi_1(t,t_1)G_{11}(t_1,t_0)dt_1\;.
\end{equation}
\end{subequations}
From Eqs. (\ref{G11})--(\ref{G21}) one can deduce
independent scalar integral equations for each component of
matrix function $\mathbf G(t,t_0)$. Indeed, after substituting
$G_{21}(t,t_0)$ from Eq. (\ref{G21}) into Eq. (\ref{G11})
the closed equation for $G_{11}(t,t_0)$ follows as
\begin{eqnarray}\label{G11final}
G_{11}(t,t_0) =\xi^{(0)}_1(t,t_0)+
\int_{t_0}^{t}\xi_1(t,t_1)G_{11}(t_1,t_0)dt_1\;.
\end{eqnarray}
In (\ref{G11final}),
\begin{eqnarray}\label{xi10}
\xi^{(0)}_1(t,t_0)=
\int_{t_0}^{t}\psi_2(t,t_1)\psi_1^{(0)}(t_1,t_0)dt_1\;
\end{eqnarray}
and
\begin{eqnarray}\label{xi1}
\xi_1(t,t_0)=
\int_{t_0}^{t}\psi_2(t,t_1)\psi_1(t_1,t_0)dt_1\;
\end{eqnarray}
is a renewal density. Analogously,
\begin{eqnarray}\label{G22final}
G_{22}(t,t_0) =\xi^{(0)}_2(t,t_0)+
\int_{t_0}^{t}\xi_2(t,t_1)G_{22}(t_1,t_0)dt_1\;,
\end{eqnarray}
where
\begin{eqnarray}\label{xi2}
\xi^{(0)}_2(t,t_0)& = &
\int_{t_0}^{t}\psi_1(t,t_1)\psi_2^{(0)}(t_1,t_0)dt_1\;, \nonumber \\
\xi_2(t,t_0) & = &
\int_{t_0}^{t}\psi_1(t,t_1)\psi_2(t_1,t_0)dt_1\;.
\end{eqnarray}
Moreover, for the off-diagonal elements of ${\mathbf G}(t,t_0)$ we find
\begin{eqnarray}\label{G12final}
G_{12}(t,t_0) & = &\psi^{(0)}_2(t,t_0)+
\int_{t_0}^{t}\xi_1(t,t_1)G_{12}(t_1,t_0)dt_1\;, \nonumber \\
G_{21}(t,t_0) & = & \psi^{(0)}_1(t,t_0)+
\int_{t_0}^{t}\xi_2(t,t_1)G_{21}(t_1,t_0)dt_1.
\end{eqnarray}

Eqs. (\ref{G11final})--(\ref{xi2}) together with Eqs. (\ref{P11})
-(\ref{P21}) present a first main result of this work.
This set of equations generalizes the
integral equations of renewal theory obtained by Cox
\cite{cox} and others \cite{boguna} to the case of non-stationary
renewal processes modulated by external signals.
The solution of the evolution operator $\mathbf \Pi(t|t_0)$
is thereby reduced
to solve the set of independent scalar integral equations for $G_{ij}(t,t_0)$.
This  presents an essential simplification as compare to the case of an
evaluation of infinite matrix integral series in Eqs.
 (\ref{structure})--(\ref{endstructure}).

\subsection{Time-homogeneous case}

In the absence of a signal,
all  two-time quantities depend only on the
time-difference $\tau=t-t_0$. In this case, the integral
equations of renewal
theory can be solved formally by use of the Laplace transform method
and the evolution operator (i. e., its Laplace-transform) can be found
explicitly. Let us  denote the Laplace transform
of any function
$F(\tau)$ below as $\tilde F(s):=\int_{0}^{\infty}\exp(-s\tau)F(\tau)d\tau$.
Then, upon Laplace transforming Eqs. (\ref{P11})--(\ref{G12final}),
using Eqs. (\ref{pdf2}), (\ref{surv2}) and some well-known theorems of
Laplace transform,
one finds
the explicit expression for the evolution operator $\tilde {\mathbf \Pi}(s)$.
It coincides with the known result
in the literature \cite{cox,boguna,goychuk03}, reading
\begin{eqnarray}
\label{stat}
\tilde {\mathbf  \Pi}(s)=\frac{1}{s}\left [ \begin{array}{cc}
 1- \frac{\tilde G(s)}{s\langle \tau_1\rangle} &
 \frac{\tilde G(s)}{s\langle \tau_2\rangle} \\
  \frac{\tilde G(s)}{s\langle \tau_1\rangle} &
  1-\frac{\tilde G(s)}{s\langle \tau_2
  \rangle}
\end{array} \right ]\;,
\end{eqnarray}
where
\begin{equation}\label{aux}
\tilde G(s)=\frac{\left(1-\tilde \psi_1(s)\right)
\left(1-\tilde \psi_2(s)\right)}{\left(1-\tilde \psi_1(s)\tilde
\psi_2(s)\right)}
\end{equation}
is an auxiliary function.

The existence of finite mean residence times $\langle
\tau_{1,2}\rangle$  implies the following useful
representation for the Laplace-transformed RTDs:
\begin{equation}\label{exp}
\tilde \psi_{1,2}(s):= 1-\langle \tau_{1,2}\rangle s\left
[1+ g_{1,2}(s)\right].
\end{equation}
In (\ref{exp}), $g_{1,2}(s)$ are corresponding functions
vanishing at $s\to 0$, i.e., $g_{1,2}(s)\to 0$. Note
that the functions  $g_{1,2}(s)$ are not necessarily
analytical. For example, $g(s)\sim s^{\gamma}$
with some real-valued exponent, $0<\gamma<1$, is allowed,  for
an example see below in Eq. (\ref{example2}).
Such non-analytical feature leads to diverging variance of RTDs.
From the formal expression (\ref{stat}) a number of important
results follows:

\subsubsection{Stationary probabilities}

The vector of stationary probabilities $\vec p^{st}=
[p_1^{st},p_2^{st}]^T$ can be evaluated as
$\vec p^{st}=\lim_{s\to
0}\left (s \tilde {\mathbf  \Pi}(s)\vec p(0) \right)$.  With
Eqs. (\ref{stat})-(\ref{exp}) one readily obtains the result
\begin{equation}\label{statpop}
p_1^{st}=\frac{\langle \tau_1\rangle}{\langle \tau_1\rangle+\langle
\tau_2\rangle},\;\;p_2^{st}=
\frac{\langle \tau_2\rangle}{\langle \tau_1\rangle+\langle
\tau_2\rangle}\;.
\end{equation}

\subsubsection{Relaxation function}
The generally non-exponential relaxation of
$\langle x(t)\rangle=x_1p_1(t)+x_2p_2(t)$ to the
stationary mean value $x_{st}=x_1p_1^{st}+x_2p_2^{st}$ is described
by the relaxation function $R(\tau)$, i.e.
\begin{eqnarray}\label{relax}
p_{1,2}(t_0+\tau)=p_{1,2}^{st}+[p_{1,2}(t_0)-p_{1,2}^{st}]\;
 R(\tau)   ,
 \end{eqnarray}
where $R(\tau)$ obeys the Laplace-transform
\begin{eqnarray}\label{R-function}
\tilde R(s)=\frac{1}{s}-\left(\frac{1}{\langle \tau_1\rangle} +
\frac{1}{\langle \tau_2\rangle} \right) \frac{1}{s^2}
\tilde G(s)\;,
\end{eqnarray}
and $\tilde G(s)$ is given by Eq. (\ref{aux}). The validity of
Eqs. (\ref{relax}), (\ref{R-function})
can be easily checked upon the use of Laplace-transformed Eq. (\ref{start})
and the result in
Eqs. (\ref{stat}), (\ref{aux}) along with the normalization
condition $p_1(t_0)+p_2(t_0)=1$ and Eq. (\ref{statpop}).
It should be emphasized here
that that the relaxation function $R(t)$ for the the considered
{\it persistent} renewal process is unique, i.e. it does not depend on
$p_{1,2}(t_0)$.
This corresponds to the situation where the random process $x(t)$
has not been prepared at $t=t_0$ in a particular state $x_1$, or $x_2$,
but rather has almost relaxed to its stationary state.
In other words, a number of
alternations occurred  before $t=t_0$ and the probability $p_{1,2}(t_0)$
to measure the particular value $x_{1,2}$ of $x(t)$ at the instant
of time $t_0$
is close to its stationary value $p_{1,2}^{st}$.
This class of initial preparations, where the relaxation function
does not depend on the actual initial probabilities, is termed
the {\it time-homogeneous} preparation class.
This preparation class \cite{GHT77,HT82}
must be distinguished from strongly non-equilibrium
initial preparations, where the system is prepared, for example,
in a particular definite state, say in the state $x_1$, with the probability
one, $p_1(t_0)=1$.

\subsubsection{Stationary autocorrelation function and regression theorem}

Let us consider next the  normalized
autocorrelation function, i. e.,
\begin{eqnarray}\label{correlation}
k(\tau)=
\lim_{t\to\infty}\frac{\langle\delta x(t+\tau)\delta x(t)\rangle }
{\langle\delta x^2\rangle_{st}}
\end{eqnarray}
of the stationary fluctuations, $\delta x(t)=x(t)-x_{st}$.
In (\ref{correlation}),
\begin{equation}\label{msq}
\langle \delta x^2\rangle_{st}=(\Delta x)^2
\frac{\langle \tau_1\rangle\langle \tau_2\rangle}{(\langle \tau_1\rangle+
\langle \tau_2\rangle)^2},
\end{equation}
is the mean squared amplitude of the stationary fluctuations and
$\Delta x=x_2-x_1$ is the fluctuation amplitude.
With $\langle \delta x(t+\tau)\delta x(t)\rangle=
 \langle x(t+\tau) x(t)\rangle-\langle x\rangle_{st}^2$,
 as $t\to \infty$, and
\begin{eqnarray}\label{explicit}
\lim_{t\to\infty}\langle x(t+\tau) x(t)\rangle
=\sum_{i=1,2}\sum_{j=1,2}x_ix_j\Pi_{ij}(\tau)p_j^{st}
\end{eqnarray}
we obtain the same result as in Ref. \cite{strat}; i.e.,
\begin{eqnarray}\label{laplace-corr}
\tilde k(s)=\frac{1}{s}-\left(\frac{1}{\langle \tau_1\rangle} +
\frac{1}{\langle \tau_2\rangle}\right)\frac{1}{s^2}\tilde G(s).
\end{eqnarray}
Upon comparison of (\ref{R-function}) with (\ref{laplace-corr})
we find the following regression theorem  for these non-Markovian
two-state processes; namely
\begin{equation}\label{regression}
R(\tau)=k(\tau)\;.
\end{equation}
The regression theorem (\ref{regression}), which relates the {\it decay} of
the {\it relaxation
function}  $R(\tau)$ to the {\it decay} of
{\it stationary fluctuations}
 $k(\tau)$, presents
 a cornerstone result for the derivation
of phenomenological linear response theory for
non-Markovian SR \cite{goychuk03}.

Usually, the Laplace-transform
(\ref{laplace-corr}) cannot be inverted analytically.
If $k(t)\geq 0$ for all times $t$,
one can define the mean correlation time:
\begin{eqnarray}\label{def1}
\tau_{corr}=\int_{0}^{\infty}k(t)dt=\lim_{s\to 0}\tilde k(s).
\end{eqnarray}
Assuming finite second moments of RDTs, $\langle \tau^2_{1,2}\rangle=
\int_{0}^{\infty}\tau^2\psi_{1,2}(\tau)d\tau$ we obtain from
Eqs. (\ref{laplace-corr}) and (\ref{def1}) the
simple result
\begin{eqnarray}\label{res1a}
\tau_{corr}=R_{NM}\tau_{M},
\end{eqnarray}
where
\begin{eqnarray}\label{res1b}
\tau_{M}=\frac{\langle \tau_1\rangle \langle \tau_2\rangle}
{\langle \tau_1\rangle + \langle \tau_2\rangle}
\end{eqnarray}
is the correlation time of the Markovian process possessing the same
MRTs $\langle \tau_{1,2}\rangle$ as the considered
non-Markovian process. The coefficient
\begin{eqnarray}\label{res1c}
R_{NM}=\frac{1}{2} \left (C^2_{1} + C^2_{2} \right )
\end{eqnarray}
presents a numerical quantifier of non-Markovian effects
in terms of the coefficients of variation of the corresponding residence
time distributions; i. e.,
\begin{equation}
C_{1,2}=\frac{\sqrt{\langle \tau_{1,2}^2\rangle -\langle \tau_{1,2}\rangle^2}}
{\langle \tau_{1,2}\rangle}\;.
\end{equation}
For example, for the stretched exponential (\ref{Weibull}) the coefficient
of variation emerges as
\begin{equation}\label{cvweib}
C=\sqrt{\frac{\Gamma(1+2/a)}{\Gamma^2(1+1/a)}-1}.
\end{equation}
For the Pareto law distribution in (\ref{Pareto}) it reads
\begin{equation}\label{cvpareto}
C=\left\{ \begin{array}{r@{\quad \quad}l}
\infty, \gamma\leq 1;\;\;\\
 \sqrt{\frac{\gamma+1}{\gamma-1}}, \; \gamma > 1\;.
\end{array} \right.
\end{equation}

As a criterion for Markovian {\it vs.} non-Markovian behavior
one can propose to test the
coefficients of variation $C_{1,2}$ of the experimentally
determined RTDs $\psi_{1,2}(t)$.
In the strict Markovian case we have $C_{1}=C_{2}=1$. Large deviations of
any of the two coefficients of variation, $C_{1,2}$,
 from unity indicate the presence of
strong non-Markovian memory effects.
The proposed test-criterion appears experimentally to be more conveniently  applied
than the direct
test of the Chapman-Kolmogorov-Smoluchowski
equation  \cite{fulinski}.
For example, in the fractal model of the ion channel gating
by Liebovitch {\it et al.} the closed residence time distribution is
fitted by (\ref{Weibull}) with $a\approx 0.2$ \cite{liebovitch}.
This yields $C_{closed}\approx 15.84$. Thus, assuming that
the open residence times are exponentially distributed, i.e. $C_{open}=1$,
one obtains
$R_{NM}\approx 126$. Furthermore,
according to Ref. \cite{mercik} BK ion channels
display a closed residence time distribution following
a Pareto law with $\gamma\approx 0.24$. In such a case,
the memory effects should depict an infinite range since $\tau_{corr}=\infty$.
In both cases, the observed two-state fluctuations do
exhibit long-range temporal correlations. The gating
dynamics is  thus clearly non-Markovian within such a
two-state description.

\subsubsection{Power spectrum of fluctuations}

For the power spectrum of fluctuations; i.e.,
\begin{equation}\label{spowerdef}
S_N(\omega)=2\langle \delta x^2\rangle_{st}
\int_{0}^{\infty}k(t)\cos(\omega
t)
dt=2\langle \delta x^2\rangle_{st}{\rm Re}\left [\tilde k(i\omega)\right ],
\end{equation}
the use of Eqs. (\ref{msq}), (\ref{laplace-corr}) in (\ref{spowerdef})
yields \cite{teich,goychuk03,lindner,HT82,strat}
\begin{equation}\label{spower}
S_N(\omega)=\frac{2(\Delta x)^2}{\langle \tau_1\rangle +\langle
\tau_2\rangle}\frac{1}{\omega^2}
{\rm Re}\left [\tilde G(i\omega)\right].
\end{equation}
 It is evident that
asymptotically, in the
limit $\omega\to \infty $, the power spectrum (\ref{spower}) is
Lorentzian in the case of time-continuous RTDs \cite{teich,remark6},
\begin{equation}\label{lorentz}
S_N(\omega)\to\frac{2(\Delta x)^2}{\langle \tau_1\rangle +\langle
\tau_2\rangle}\frac{1}{\omega^2},\;\;{\rm as}\;\;\omega\to\infty \;.
\end{equation}
This follows from the fact that
$\lim_{\omega\to \infty}\psi_{1,2}(i\omega)=0$ and thus
$\lim_{\omega\to \infty}\tilde G(i\omega)=1$ \cite{remark6}. Practically this
situation occurs for $\omega\gg \langle \tau_{1,2}\rangle^{-1}$.
On the other hand,
one can deduce from Eq. (\ref{spowerdef}) that in the opposite limit
for $\omega\to 0$,
\begin{equation}\label{zero}
S_N(\omega)\to S_N(0) =2\langle \delta x^2\rangle_{st}\tau_{corr},
\end{equation}
where $\langle \delta x^2\rangle_{st}$ is the mean-squared
amplitude of stationary fluctuations given by (\ref{msq})
and $\tau_{corr}$ is given in  Eq. (\ref{res1a}). A very
interesting situation emerges for $\tau_{corr}\to
\infty$, implying $S_N(0)\to \infty$.
This occurs when at least one of the residence time
distributions possesses a diverging variance, cf. Eqs.
(\ref{res1a})-(\ref{res1c}). In such a case, for
the low-frequency region $\omega<\langle \tau_{1,2}\rangle ^{-1}$
the power spectrum drastically
differs from the Lorentzian form.
For example, for a symmetric
TSRP with the survival probabilities given by the Pareto distributions
(\ref{Pareto}) one can show \cite{teich} (see also below) that for
$0<\gamma<1$, $S_N(\omega)\sim
1/\omega^{1-\gamma}$. For $\gamma\to 0$ this corresponds to
celebrated $1/f$
noise \cite{teich,remark7}.

\section{Phenomenological theory of linear response}

\begin{figure}
\epsfig{file=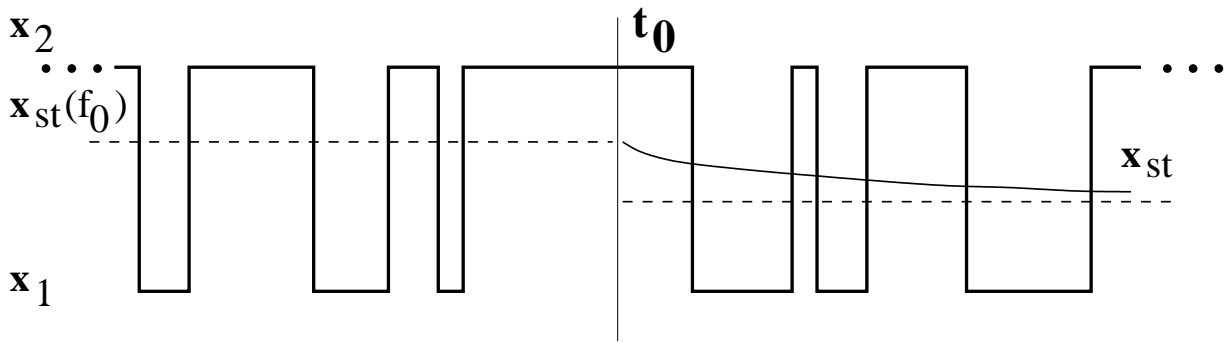,width=.9\textwidth}

\caption{Relaxation of a perturbed persistent renewal process $x(t)$.
A constant force $f_0$ is applied long before and is released at $t=t_0$.
The mean value $\langle x(t)\rangle$ relaxes from the constrained
stationary value $x_{st}(f_0)$ to its true stationary  value
$x_{st}$.
 }
\label{Fig1}
\end{figure}

It is possible to predict the linear response of the underlying stochastic
process $x(t)$ to the external driving $f(t)$ by referring only to information
on its stationary properties,i.e., without explicit knowledge of
the concrete mechanism at work
by which the process $x(t)$ is perturbed by the external signal.
The phenomenological theory of
linear response  for general stochastic processes \cite{HT82,thesisPH} and
for thermal physical systems \cite{kubo} provides a
very useful and widely applied tool to answer this question.
It is also the only method available if no further detailed
knowledge of the microscopic dynamics is at hand for the {\it observed}
two-state dynamics. This is the common experimental situation.
The common linear response approximation
  \begin{eqnarray}\label{response2}
\langle \delta x(t)\rangle:=
\langle x(t)\rangle - x_{st}
=\int_{-\infty}^{t}\chi(t-t')f(t')dt',
\end{eqnarray}
holds independently of the underlying stochastic dynamics \cite{HT82}.
In (\ref{response2}),
$\chi(t)$ denotes the linear response function in the time domain.
The universality of the relation (\ref{response2}) allows one
to find the linear response function $\chi(t)$ using a properly designed form
of the perturbation $f(t)$.
Within the phenomenological approach
it can
be obtained following  an established procedure \cite{kubo}:
(i) First, apply a small
static ``force'' $f_0$, (ii) then,
let the process $x(t)$ relax to the constrained
stationary state with mean value $x_{st}(f_0)$, and finally (iii)
suddenly remove the ``force''
at $t=t_0$, see Fig. \ref{Fig1}.

Then, in accord with (\ref{response2}) the response function reads
\begin{eqnarray}\label{response3}
\chi(\tau)=-\frac{1}{f_0}\frac{d}{d\tau}
\langle\delta x(t_0+\tau)\rangle, \;\;\;\;\;\tau>0,
\end{eqnarray}
where $\langle\delta x(t_0+\tau)\rangle=x_1p_1(t_0+\tau)+x_2p_2(t_0+\tau)$
is determined by  (\ref{relax}) with the initial $p_{1,2}(t_0)$ taken as
$p_{1,2}(t_0)=\langle \tau_{1,2}(f_0) \rangle/[\langle
\tau_{1}(f_0)\rangle+\langle \tau_{2}(f_0)\rangle]$.
The limit $f_0\to 0$ is implicitly assumed in Eq. (\ref{response3}).
Expanding
$p_{1,2}(t_0)$ to first order in $f_0$ we find with $\Delta x=x_2-x_1$
\begin{eqnarray}\label{expand}
\langle\delta x(t_0+\tau)\rangle =
\frac{\langle \delta x^2\rangle_{st}}{\Delta x}
[\beta_2-\beta_1] R(\tau) f_0 +o(f_0),
\end{eqnarray}
where
\begin{eqnarray}\label{beta}
\beta_{1,2}:=\frac{d\ln \langle \tau_{1,2}(f_0)\rangle}{df_0}|_{f_0=0}\;\;.
\end{eqnarray}
Note that in the derivation of this result it is tacitly assumed
that the initial constrained stationary populations $p_{1,2}(t_0)$
at $t=t_0$ belongs to the class of time-homogeneous initial
preparations \cite{HT82}
for the process $x(t)$ in the absence of applied force. This seems a
natural and intuitively clear  assumption in view of the facts that
the limit
$f_0\to 0$ has to be taken
in (\ref{response3}) at the very end of calculation, and the considered
process is persistent. Nevertheless,
this commonly accepted assumption
is a hidden hypothesis which, strictly speaking,
cannot be
proven within the phenomenological approach.

Upon combining (\ref{expand})  with the regression
theorem (\ref{regression}) we obtain from (\ref{response3}),
after taking the limit $f_0\to 0$,
the {\it fluctuation theorem} \cite{goychuk03}
\begin{eqnarray}\label{fluctuation1}
\chi(\tau)=-[\beta_2-\beta_1]\frac{\theta(\tau)}{\Delta x}
\frac{d}{d\tau} \langle\delta x(t+\tau)\delta x(t)\rangle_{st} \;,
\end{eqnarray}
wherein $\theta(t)$ denotes here the unit step function. The non-Markovian
fluctuation theorem
(\ref{fluctuation1}) presents a prominent result \cite{goychuk03};
in particular, it does not assume thermal equilibrium \cite{HT82}.
In the frequency domain it reads
\begin{eqnarray}\label{FTfreq1}
\tilde \chi (\omega) & = &\frac{(\beta_2-\beta_1)\langle
\delta x^2\rangle_{st} }{\Delta x}\Big [ 1+ i\omega \tilde k (-i\omega)\Big],
\end{eqnarray}
where $
\tilde\chi(\omega)=\int_{-\infty}^{\infty}
\chi(t){\rm e}^{i\omega t}dt$ denotes the linear response function
in the frequency domain, and $\tilde k(s)$ is given by Eq. (\ref{laplace-corr}).
Substitution of Eqs. (\ref{laplace-corr}) and (\ref{msq}) in (\ref{FTfreq1})
yields
 \begin{eqnarray}\label{FTfreq2}
\tilde \chi (\omega) & = &\frac{(\beta_2-\beta_1)\Delta x }
{\langle \tau_1\rangle +\langle \tau_2 \rangle}\frac{i}{\omega}\tilde
G(-i\omega),\;
\end{eqnarray}
where $\tilde G(s)$ is given in (\ref{aux}). The expression (\ref{FTfreq2})
together with Eq. (\ref{aux})
connects the linear response function $\tilde \chi(\omega)$
with the Laplace-transformed residence time distributions
$\tilde \psi_{1,2}(i\omega)$, i.e.,  with the characteristic
functions of the RTDs.

If, in addition, the mean residence times  obey
the thermal detailed balance relation
\begin{eqnarray}\label{detbal}
\frac{\langle \tau_{1}(f_0)\rangle}{\langle \tau_{2}(f_0)\rangle}
=\exp\Big(\frac{-\epsilon(T)-f_0\Delta x }{k_BT}\Big),
\end{eqnarray}
where $\epsilon(T)$ is the free-energy difference between two  metastable
states, we recover for the fluctuation theorem in (\ref{fluctuation1}) the
 form that characterizes classical  equilibrium
dynamics \cite{HT82,kubo,new2}; i.e.,
\begin{eqnarray}\label{equilibrium}
\chi(\tau)=-\frac{\theta(\tau)}{k_BT}
\frac{d}{d\tau} \langle\delta
x(\tau)\delta x(0)\rangle_{st}.
\end{eqnarray}
Eq. (\ref{FTfreq2}) then yields
\begin{eqnarray}\label{FTfreq3}
\tilde \chi (\omega) & = &\frac{(\Delta x)^2 }{k_BT}\frac{1}
{\langle \tau_1\rangle +\langle \tau_2 \rangle}\frac{i}{\omega}\tilde
G(-i\omega)\;.
\end{eqnarray}
For example,
this result is valid for an Arrhenius-like
dependence of $\langle \tau_{1,2}\rangle$ on
temperature $T$ and force $f_0$; i.e.
\begin{eqnarray}\label{Arrhenius}
\langle \tau_{1,2}(f_0)\rangle & = &A_{1,2} \exp\Big (\frac{
\Delta U_{1,2}\mp \Delta x_{1,2} f_0}{k_BT}  \Big),
\end{eqnarray}
where $\Delta U_{1,2}$ are the heights of activation barriers,
$\Delta x_1=z\Delta x$, $\Delta x_2=(1-z)\Delta x$ with $\Delta x=x_2-x_1,
\;0<z<1$.
Eq. (\ref{equilibrium}) presents a key result because it provides
a link between
the phenomenological theory of linear response theory and the actual physical
processes which are in {\it thermal equilibrium} and do
exhibit long-range time correlations. Let us assume, for example, the following situation:
The observed two-state process results from thermally activated
transitions in a complex potential energy landscape $U(\vec x)$
possessing two domains of attraction (i.e., two metastable states) separated
by distance $\Delta x$ along the direction of the reaction
coordinate $x$
which describes transitions
between the metastable states. Next, let us assume that the coupling
of the external force $f(t)$ to the dynamics has the potential
energy form $U_{int}=-xf(t)$.
Then, the classical equilibrium fluctuation
theorem (\ref{equilibrium}) follows  from
 first principles \cite{kubo}, or, likewise, from a mesoscopic starting point
in terms of the generalized master equation for the thermal equilibrium dynamics \cite{new2};
in other words, it is exact.
The non-exponential features of the RTDs in the described situation stems
from the motions ``perpendicular'' to the above reaction coordinate $x$.
In such a case, the thermodynamic relations like
Eq. (\ref{detbal}) are compatible
with non-Markovian kinetics. This is the case where
the phenomenological theory of linear response in non-Markovian
systems  has a firm foundation. The readers should be warned,
however, that the phenomenological theory is not universally valid
for  nonequilibrium physical systems. Nevertheless, below we explicitly
define an  universality class of such systems
(which are beyond the thermal equilibrium class)
where its validity can be proven on a more general basis.

\section{Asymptotic response theory based on driven
renewal equations}

Starting from the driven renewal equations (\ref{P11})-(\ref{G12final})
one can develop the theory of the linear and the
nonlinear response which possesses a broader range of validity
as compared to the above phenomenological theory.
For a periodic signal (switched on in the infinite past)
like in (\ref{signal}),
the conditional survival probabilities
$\Phi_{1,2}(\tau|t):=\Phi_{1,2}(t+\tau,t)$
 acquire (at asymptotic times $t\gg t_0$)
 the time periodicity in $t$ of the driving signal and therefore can be
expanded into the Fourier series, i.e.,
\begin{equation}\label{fourier}
\Phi_{1,2}(\tau|t)=\sum_{n=-\infty}^{\infty}\Phi^{(n)}_{1,2}(\tau)
\exp[-in\Omega t],\;\; \Phi^{(-n)}_{1,2}(\tau)=
\left[\Phi^{(n)}_{1,2}(\tau)  \right]^*.
\end{equation}
Similar expansions hold also  for the conditional residence time
distributions $\psi_{1,2}(\tau|t)$ with the corresponding
 expansion coefficients
$\psi^{(n)}_{1,2}(\tau)=-\frac{d}{d\tau}\Phi^{(n)}_{1,2}(\tau)$.
Note that $\Phi^{(0)}_{1,2}(\tau)$ and $\psi^{(0)}_{1,2}(\tau)$
in this section denote the Fourier expansion coefficients with $n=0$.
These quantities
are clearly not related to the survival functions (\ref{surv2})
and RTDs (\ref{pdf2})  of the
first time interval.
We hope that such use of notations will not confuse the readers.
The corresponding Laplace-transformed quantities of
the $\tau$-dependent Fourier coefficients
$\tilde \psi^{(n)}_{1,2}(s)$ and $\tilde \Phi^{(n)}_{1,2}(s)$
in (\ref{fourier})
are related by
\begin{equation}\label{connect}
\tilde \psi^{(n)}_{1,2}(s)=\delta_{n,0}-s\tilde \Phi^{(n)}_{1,2}(s).
\end{equation}
Our goal is to evaluate the asymptotic behavior of the populations
$p_{1,2}^{(as)}(t)$ and of the mean value $\langle x^{(as)}(t)\rangle$.
To do so, one
needs to determine the asymptotic evolution operator ${\mathbf \Pi^{(as)}(t)}:=
\lim_{t_0\to-\infty}{\mathbf \Pi(t|t_0)}$. Obviously, $\Pi^{(as)}_{11}(t)=
\Pi^{(as)}_{12}(t)$ and $\Pi^{(as)}_{22}(t)=
\Pi^{(as)}_{21}(t)$. Moreover, $p_{1}^{(as)}(t)=\Pi^{(as)}_{11}(t)$,
$p_{2}^{(as)}(t)=\Pi^{(as)}_{22}(t)$. Next, let us
define the auxiliary quantity ${\mathbf G^{(as)}(t)}$ as
${\mathbf G^{(as)}(t)}:=\lim_{t_0\to-\infty}
{\mathbf G}(t,t_0)$. Then,
Eqs. (\ref{P11}), (\ref{G11final}) in the
limit $t_0\to -\infty$ yield
\begin{eqnarray}\label{asymp1}
p_1^{(as)}(t) & = &
\int_{-\infty}^{t}\Phi_1(t,t_1)G_{11}^{(as)}(t_1)dt_1,
\end{eqnarray}
where $G_{11}^{(as)}(t)$ is solution of the integral equation:
\begin{eqnarray}\label{asymp2}
G_{11}^{(as)}(t) & = &\int_{-\infty}^{t}\xi_1(t,t_1)
G_{11}^{(as)}(t_1)dt_1,
\end{eqnarray}
with the renewal density $\xi_1(t,t_1)$ given in Eq. (\ref{xi1}).
The equation
determining $p_2^{(as)}(t)$ likewise reads
\begin{eqnarray}\label{asymp3}
p_2^{(as)}(t) & = &
\int_{-\infty}^{t}\Phi_2(t,t_1)G_{22}^{(as)}(t_1)dt_1,
\end{eqnarray}
where $G_{22}^{(as)}(t)$ is the solution of integral equation
\begin{eqnarray}\label{asymp4}
G_{22}^{(as)}(t) & = &\int_{-\infty}^{t}\xi_2(t,t_1)
G_{22}^{(as)}(t_1)dt_1,
\end{eqnarray}
with $\xi_2(t,t_1)$ given in Eq. (\ref{xi2}). Note that the
conditional renewal densities $\xi_{1,2}(\tau|t):=\xi_{1,2}(t+\tau,t)$
also acquire a time-periodicity in $t$
and can be represented in the form like (\ref{fourier}) with the
corresponding expansion coefficients $\xi_{1,2}^{(n)}(\tau)$.
One can show that
the corresponding Laplace-transformed quantities
$\tilde \xi_{1,2}^{(n)}(s)$ are related with the quantities
$\tilde \psi_{1,2}^{(n)}(s)$ as follows:
\begin{eqnarray}\label{rel}
\tilde \xi_1^{(n)}(s)&=&\sum_{m=-\infty}^{\infty}\tilde \psi_2^{(m)}(s)
\tilde \psi_1^{(n-m)}(s+im\Omega), \nonumber \\
\tilde \xi_2^{(n)}(s)&=&\sum_{m=-\infty}^{\infty}\tilde \psi_1^{(m)}(s)
\tilde \psi_2^{(n-m)}(s+im\Omega)\;.
\end{eqnarray}
For periodic driving $f(t)$, both $p_{1,2}^{(as)}(t)$ and $G_{11,22}^{(as)}(t)$
must be periodic functions of time \cite{review1} and can be expanded into
Fourier series:
\begin{equation}\label{pk}
p_1^{(as)}(t)=\sum_{k=-\infty}^{\infty} p_{1,2}^{(k)} e^{-i k\Omega t},\;
p_{1,2}^{(-k)}=[p_{1,2}^{(k)}]^*
\end{equation}
and
\begin{equation}\label{gk}
G_{11,22}^{(as)}(t)=\sum_{k=-\infty}^{\infty} g_{1,2}^{(k)}
 e^{-i k\Omega t},\;
g_{1,2}^{(-k)}=[g_{1,2}^{(k)}]^*,
\end{equation}
respectively.

Using Eqs. (\ref{signal}), (\ref{response2}) and the expansion
(\ref{pk}) one can show that
the coefficient $p_1^{(1)}$ in Eq. (\ref{pk}) determines the
{\it linear response function} $\tilde \chi(\Omega)$ in the frequency domain
as
\begin{eqnarray}\label{proced1}
\tilde \chi(\Omega) =-\frac{2\Delta x}{f_0} p_1^{(1)}
\end{eqnarray}
in the limit $f_0\to 0$.
Moreover, from the normalization
condition $p_1^{(as)}(t)+p_2^{(as)}(t)=1$ it follows that
\begin{eqnarray}\label{norm}
p_1^{(0)}+p_2^{(0)}=1,\;\;\; p_1^{(n)}=-p_2^{(n)}\;\;{\rm for}\;\; n\neq 0.
\end{eqnarray}
Upon substituting Eqs. (\ref{pk}) and (\ref{gk}) and the expansions like
Eq. (\ref{fourier}) into Eqs.
(\ref{asymp1})-(\ref{asymp4}), performing the time integration and
comparing the coefficients of the Fourier expansions on the left and right
hand sides of the corresponding equations we finally end up with:
\begin{eqnarray}\label{fin1}
p_1^{(k)} & = &\sum_{n=-\infty}^{\infty}\tilde
\Phi_1^{(n)}(-ik\Omega)g^{(k-n)}_1,\\ \label{fin2}
g_1^{(k)} &= &\sum_{n=-\infty}^{\infty}\sum_{m=-\infty}^{\infty}
\tilde \psi_2^{(m)}(-ik\Omega)\tilde \psi_1^{(n-m)}
 (-i[k-m]\Omega )g^{(k-n)}_1\;,
\end{eqnarray}
and
\begin{eqnarray}\label{fin3}
p_2^{(k)} & = &\sum_{n=-\infty}^{\infty}\tilde
\Phi_2^{(n)}(-ik\Omega)g^{(k-n)}_2,\\ \label{fin4}
g_2^{(k)} &= &\sum_{n=-\infty}^{\infty}\sum_{m=-\infty}^{\infty}
\tilde \psi_1^{(m)}(-ik\Omega)\tilde \psi_2^{(n-m)}
 (-i[k-m]\Omega )g^{(k-n)}_2\;.
\end{eqnarray}
The relations (\ref{fin1})-(\ref{fin4}) also serve as the basis for a
response theory without restriction on the linear response
approximation. In order to apply these equations, one has to
specify the expansion coefficients in Eq. (\ref{fourier}), i.e. to
concretize the way how the external signal $f(t)$ enters the
conditional residence time
distributions $\psi_{1,2}(\tau|t)$, or, equivalently, the
conditional survival
probabilities $\Phi_{1,2}(\tau|t)$ to the required order in the
signal amplitude $f_0$.
 It is worth to notice that the solutions of
Eqs. (\ref{fin2}), (\ref{fin4}) are defined up to some arbitrary constants
which can be fixed at the end of calculations by applying the
normalization relations in (\ref{norm}).

In the linear response
approximation, $\tilde \Phi_{1,2}^{(0)}(s)=\tilde \Phi_{1,2}(s)$, i.e.,
$\tilde \Phi_{1,2}^{(0)}(s)$ coincide with the unperturbed survival
probabilities $\tilde \Phi_{1,2}(s)$.
Moreover,
$\tilde \Phi_{1,2}^{(1)}(s)\propto f_0$. All the higher order terms
$\tilde \Phi_{1,2}^{(n\geq 2)}(s)$ can be neglected,
being of higher order proportional to $f_0^n,\;n\geq 2$.
 The same holds true for $\tilde \psi_{1,2}^{(n)}(s)$.
After some cumbersome algebra,  one finds 
from Eqs. (\ref{fin1})-(\ref{fin4})
an expression for $p_1^{(1)}$, which then by use of relation
(\ref{proced1}) yields
\begin{eqnarray}\label{mainres}
\tilde \chi(\Omega)=-\frac{2i\Delta x}{f_0\Omega}\frac{1}{\langle \tau_1\rangle
+\langle \tau_2\rangle}\frac{\tilde \psi_2^{(1)}(-i\Omega)
[1-\tilde \psi_1(-i\Omega)]-\tilde \psi_1^{(1)}(-i\Omega)
[1-\tilde \psi_2(-i\Omega)]}{1-\tilde\psi_1(-i\Omega)\tilde\psi_2(-i\Omega)}.
\end{eqnarray}
The result in Eq. (\ref{mainres}) presents a second
cornerstone result of this work.
Note that this  general result depends on the quantities
$\tilde \psi^{(1)}_{1,2}(s)\propto f_0$ which do not follow directly from
the characteristic functions of stationary RTDs, i.e.,
$\tilde \psi_{1,2}(s)$, but their knowledge requires
one to specify a microscopic model. Generally, Eq. (\ref{mainres})
is not mathematically reducible to the result (\ref{FTfreq2})
of the phenomenological theory. A question arises whether such a reduction
is possible in practice and the phenomenological theory of linear
response can be put on a more firm ground beyond the
time-homogeneous preparation class result
in Eq. (\ref{FTfreq2}) of which the thermal equilibrium
result in Eq. (\ref{FTfreq3}) is a special case.
Below we describe a rather broad class of
relevant systems.

\subsection{Models with form-invariant RTDs}

Let us assume that the survival probability  and the corresponding
RTD can be parameterized by a single frequency parameter $\nu$
with has the meaning of an inverse mean residence time, i.e., $\nu=\langle
\tau \rangle^{-1}$.
Furthermore, we assume that a weak signal $f(t)$ causes $\nu$ to became
time-dependent, i. e.,
\begin{equation}\label{nut}
\nu\to\nu(t)=\nu[1-\beta f(t)]\;,
\end{equation}
with $\beta \ll 1/f_0$ (the subscripts $1,2$ are suppressed).
Moreover, the survival probabilities become modified
applying the following rule:
$\nu\tau\to \int_{t}^{t+\tau}\nu(t')dt'$. More generally,
let us consider arbitrary survival probabilities of the
form (\ref{gen}) generalized to the time-inhomogeneous case
in the following way
\begin{eqnarray}\label{genmod}
\Phi(\tau)\to \Phi(\tau|t)=
\sum_{i=1}^{\infty}c_i\exp\Big (-\int_{t}^{t+\tau}
\nu_i(t')dt'\Big ),\;\sum_i c_i=1.
\end{eqnarray}
In (\ref{genmod}), we assume that (to leading order) neither the
expansion coefficients $c_i$, nor the ratios between any of $\nu_i(t)$
and $\nu_j(t)$ are modified by the applied signal $f(t)$, i.e.,
\begin{equation}\label{scale}
\frac{\nu_i(t)}{\nu_j(t)}=a_{ij} \;,
\end{equation}
with $a_{ij}$ being some structural constants.
This covers fractal (although {\it not} multi-fractal) time distributions.
Put differently, the scaling law by which the whole hierarchy
of rate constants is produced from a single rate constant is invariant
of the applied signal. If the mean residence time
$\langle \tau \rangle=\sum_i c_i/\nu_i$
exists, one can always set $\nu=\langle \tau\rangle^{-1}$
as the relevant rate constant in the absence of driving. This rate
will acquire
an explicit time-dependence like in (\ref{nut}) when the signal is switched on.
 Given our assumptions, all the time-dependent rates
$\nu_i(t)$ in (\ref{genmod})
 will
 be proportional to the  rate $\nu(t)$ in Eq. (\ref{nut}). Then,
 in the lowest first order in $\beta f_0$, we find
\begin{eqnarray}\label{1order}
\Phi(\tau|t)=\Phi(\tau)+\beta\psi(\tau)\int_{t}^{t+\tau}f(t')dt'\;.
\end{eqnarray}
From (\ref{1order})  we obtain
upon observing Eq. (\ref{signal})
\begin{eqnarray}\label{fi1}
\Phi_{1,2}^{(1)}(\tau)=\frac{1}{2}i\frac{\beta_{1,2}f_0}{\Omega}
\psi_{1,2}(\tau)\Big [\exp(-i\Omega \tau) -1 \Big]
\end{eqnarray}
and
\begin{eqnarray}\label{fi1s}
\tilde \Phi_{1,2}^{(1)}(s)=-\frac{1}{2}i\frac{\beta_{1,2}f_0}{\Omega}
\Big [\tilde \psi_{1,2}(s)- \tilde \psi_{1,2}(s+i\Omega) \Big]\;.
\end{eqnarray}
Observing Eq. (\ref{connect}) by taking into account
$\tilde\psi_{1,2} (0)=1$ in Eq. (\ref{fi1s}) thus yields
\begin{eqnarray}\label{class}
\tilde \psi_{1,2}^{(1)}(-i\Omega)=-\frac{1}{2}\beta_{1,2}
f_0[1-\tilde \psi_{1,2}(-i\Omega)]\;.
\end{eqnarray}
Substituting (\ref{class}) into (\ref{mainres}) we recover
the result of the phenomenological theory in
Eq. (\ref{FTfreq2}). In conclusion, for the considered class of models
the non-equilibrium fluctuation theorem (\ref{FTfreq2}) is well
justified. This model class can therefore be reconciled with
the assumption of time-homogeneous initial preparations used
in the phenomenological theory of linear
response (see Sec. III). This assumption is naturally not always
justified a priori. It  rather delimits an important and rather
broad class of corresponding physical systems.
Nevertheless, the equilibrium fluctuation theorem (\ref{equilibrium})
presents a fundamental
relation which must be obeyed for all thermal equilibrium systems. This
imposes a salient restriction on mesoscopic models
leading to the observed equilibrium non-Markovian dynamics.
In particular, if one knows that the considered system is in the
thermal equilibrium, one must use the
rigorous relation (\ref{FTfreq3}), rather than (\ref{mainres})
for the calculation of the linear response. This
constitutes the essence of the phenomenological theory of
non-Markovian stochastic resonance developed in Ref. \cite{goychuk03}.
For other systems, e.g., for those modelling  neuronal dynamics
(which are far away from thermal equilibrium) the use of Eq. (\ref{mainres})
is preferred. In order to apply Eq. (\ref{mainres}), however, one must
also specify the
underlying non-equilibrium microscopic dynamics in the presence of
a time-periodic stimulus. This means that the time-inhomogeneous
conditional RTDs
$\psi_{1,2}(\tau|t)$ must be measured, or modelled (to the
linear order) in the driving signal strength.
We next present a detailed study of non-Markovian Stochastic Resonance
in {\it thermal equilibrium} systems that do exhibit prominent temporal
long-range time correlations \cite{goychuk03}.

\section{Stochastic resonance}

In the presence of applied periodic
signal (\ref{signal}), the  spectral power amplification
(SPA) \cite{review1,new3},
$\eta(\Omega)=|\tilde\chi(\Omega)|^2$
reads by use of the fluctuation theorem
in (\ref{FTfreq3}) upon combining
  (\ref{correlation}), (\ref{laplace-corr}),
(\ref{msq}), (\ref{Arrhenius}) as follows
\begin{equation}\label{res1}
\eta(\Omega,T)=
\frac{(\Delta x/2)^4}{(k_BT)^2}
\frac{\nu^2(T)}{\cosh^4\left[\epsilon(T)/(2k_BT)\right]}
\frac{|\tilde G(i\Omega)|^2}{\Omega^2}.
\end{equation}
In (\ref{res1}),
$\nu(T)=\langle \tau_1\rangle^{-1}+\langle \tau_2\rangle^{-1}$
denotes the sum of effective rates.  The quantity
$\epsilon(T)=\Delta U_2-\Delta U_1+T\Delta S$ denotes the
free-energy difference
between the metastable states which includes the entropy difference
$\Delta S:=S_2-S_1=k_B\ln(A_2/A_1)$.
In the Markovian case we obtain $\tilde G(s)=s/(s+\nu)$ and  (\ref{res1})
equals the known result, see in \cite{review1,new3}.

The signal-to-noise ratio (SNR) is given within linear response
approximation by
\begin{eqnarray}
{\rm SNR}(\Omega,T):=\frac{\pi f_0^2|\tilde\chi(\Omega)|^2}{S_N(\Omega)},
\end{eqnarray}
where $S_N(\omega)$, Eq. (\ref{spowerdef}), is the spectral power of
stationary fluctuations \cite{review1}.
 By use of  (\ref{res1}), we obtain
\begin{eqnarray}\label{res3}
{\rm SNR}(\Omega,T)=\frac{\pi f_0^2(\Delta x/2)^2}{2(k_BT)^2}\frac{\nu(T)}
{\cosh^2\left[\frac{\epsilon(T)}{2k_BT)}\right]}\; N(\Omega),
\end{eqnarray}
where the term
\begin{eqnarray}\label{N}
N(\Omega)=\frac{
|\tilde G(i\Omega)|^2}{{\rm Re}[\tilde G(i\Omega)]}
\end{eqnarray}
denotes a frequency- and temperature-dependent non-Markovian correction.
For arbitrary continuous $\psi_{1,2}(\tau)$ the function $N(\Omega)$
approaches unity for high-frequency signals,
$\Omega\gg \langle \tau_{1,2}\rangle^{-1}$.
Thus, Eq. (\ref{res3}) reduces in this limit
to the known Markovian result \cite{review1},
 i.e. the Markovian limit of SNR is assumed asymptotically in the high
 frequency regime.
More interesting, however, is the result for small frequency driving.
In the zero-frequency limit we find
$N(0)=1/R_{NM}$ with $R_{NM}$ given in Eq. (\ref{res1c}).
With $R_{NM}=\infty$ as it is the case for the Pareto distribution
(\ref{Pareto}) with $0<\gamma<1$,
$N(0)=0$; i.e. ${\rm SNR}(\Omega=0,T)=0$ as well.
Consequently, ultra-slow signals are
difficult to detect within the SNR-measure in a strongly
non-Markovian situation.

\subsection{Symmetric SR}

As a first example, we address  non-Markovian SR in a symmetric system
with the
survival probabilities $\Phi_{1,2}(\tau)$ described by the
identical power
laws (\ref{Pareto}) with $\nu=\langle \tau\rangle^{-1}$ determined
from Eq. (\ref{rate-t}) with $f(t)=0$, $\nu_{1,2}^{(0)}=\nu_0$
and $\Delta U_{1,2}=\Delta U$.
In this case, the Laplace-transformed  RTDs read
\begin{eqnarray}\label{example1}
\tilde \psi(s)=1-\Big (\gamma \langle \tau \rangle s\Big )^{\gamma+1}
\exp(\gamma\langle \tau \rangle s)
\Gamma\Big (-\gamma,\gamma \langle \tau \rangle s\Big),
\end{eqnarray}
where $\Gamma(x,y)$ is the incomplete gamma-function
\cite{stegun}. For $0<\gamma<1$, the distribution (\ref{example1})
has a diverging variance; its small-$s$ expansion
reads
\begin{eqnarray}\label{example2}
\tilde \psi(s)\approx 1-\langle \tau\rangle
s+\gamma^{\gamma}\Gamma(1-\gamma)[\langle \tau\rangle
s]^{1+\gamma}.
\end{eqnarray}

Using (\ref{example2}) and
(\ref{aux}) in (\ref{spower}) we obtain for the
low-frequency part of the power
spectrum
\begin{eqnarray}\label{example3}
S_N(\omega)\approx \frac{1}{2}(\Delta x)^2
\Gamma(1-\gamma)\sin(\pi\gamma/2)
\frac{\gamma\langle \tau\rangle}{[\gamma\langle \tau\rangle
\omega]^{1-\gamma}}.
\end{eqnarray}
To obtain the spectral amplification
(\ref{res1}) and the SNR (\ref{res3}) numerically
one has to use Eq. (\ref{example1}) in Eq. (\ref{aux}).
For $\gamma>1$, the power spectrum of this process mimics
a conventional Lorentzian. Moreover, for $\gamma\gg 1$,
$C\approx 1$, cf. (\ref{cvpareto}). Thus, one can expect that for large
$\gamma$ the considered situation does not differ much from the Markovian
case, at least qualitatively. Indeed,  for  very large
$\gamma\sim 100$ the
discrepancy with the Markovian case in the SNR behavior versus noise
intensity $D=k_BT$ is
not detectable. The well known bell-shaped Stochastic Resonance
behavior is reproduced with the maximum at $D=\Delta U/2$. Nevertheless,
in the behavior of  $\eta(\Omega,T)$ some discrepancy
still remains detectable even for such large $\gamma$ (not shown).

Next, the case with $0<\gamma\leq 1$ is of major interest as
it is qualitatively very distinct from
the Markovian Stochastic Resonance, see Fig. 2. The reason is that
the mean correlation time $\tau_{corr}$ in Eq. (\ref{def1}) becomes
formally infinite and the power spectrum exhibits a typical
$1/f^{\alpha}$-characteristics, with $\alpha=1-\gamma$, cf. Eq. (\ref{example3}).
Nevertheless, an important time scale of the stochastic dynamics
does still exist: It is defined by the
mean time of stochastic turnovers between the metastable states, $\tau_{0}(D)=2
\langle \tau\rangle$. Invoking the reasoning of a stochastic
synchronization of stochastic resonance \cite{new3}
one can expect Stochastic Resonance to occur
when the time scale of
stochastic turnovers $\tau_0(D)$  matches
the period of external driving ${\cal T}=2\pi/\Omega$, i.e.,
$\tau_0(D)\sim {\cal T}$.
Indeed, Fig. 2 (a) unambiguously demonstrates  the Stochastic Resonance
phenomenon for
a non-Markovian system  with $\gamma=0.2$.
Thus, the interpretation
of SR as the  phenomenon caused by stochastic synchronization
between the time-scales of the random, temperature driven transitions
and the external periodic modulations \cite{review1,new3}
can be extended
even onto this
extreme non-Markovian case (with diverging mean correlation time,
$\tau_{corr}=\infty$).
 Note, however, that the maximal value of the spectral
amplification of signal is strongly suppressed in the present case
by the factor of about $20$
as compare
with the corresponding Markovian counterpart possessing the same
$\langle \tau\rangle$, see Fig. 2(b).

\begin{figure}

\epsfig{file=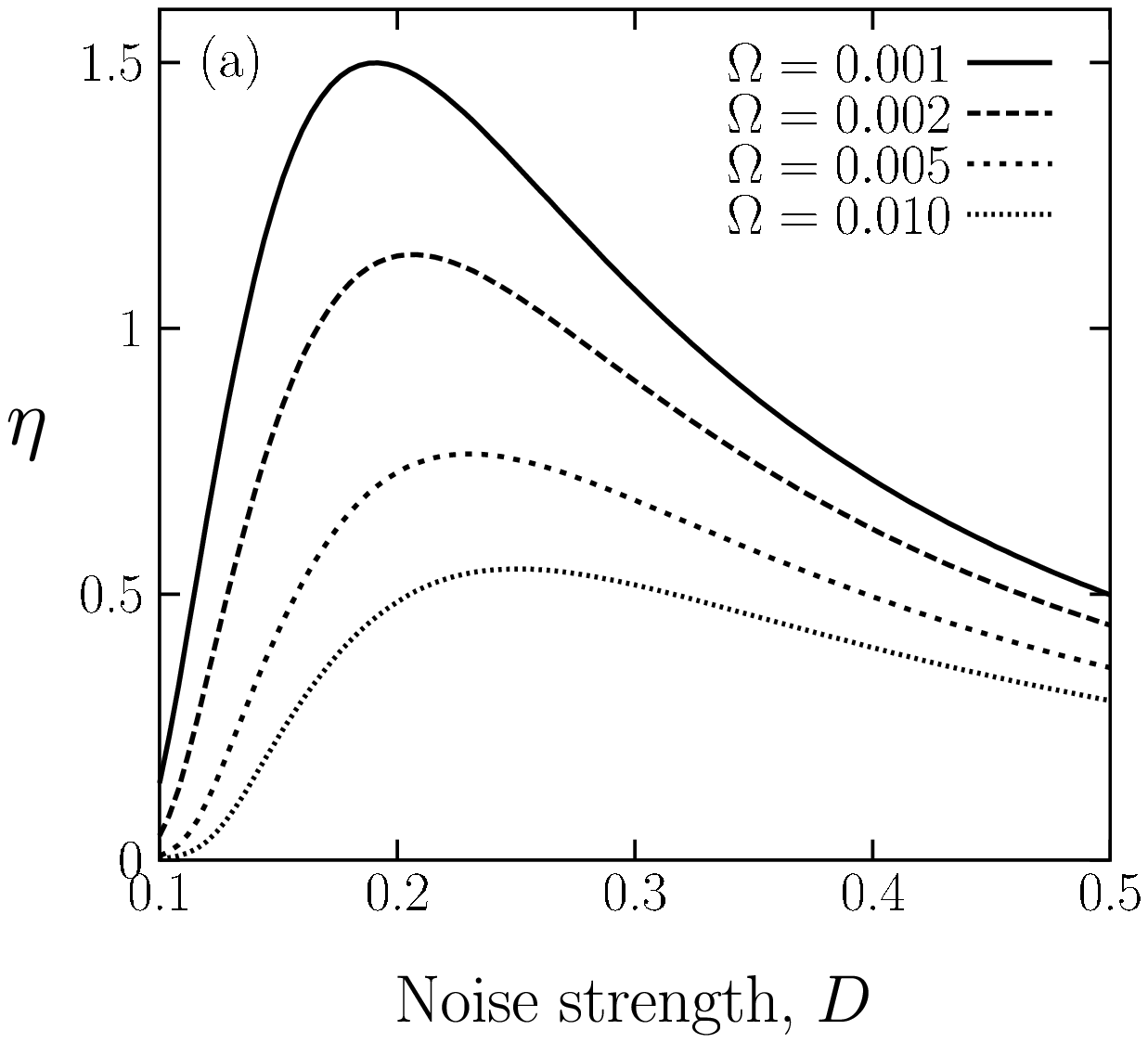,width=0.49\textwidth}
\hfill
\epsfig{file=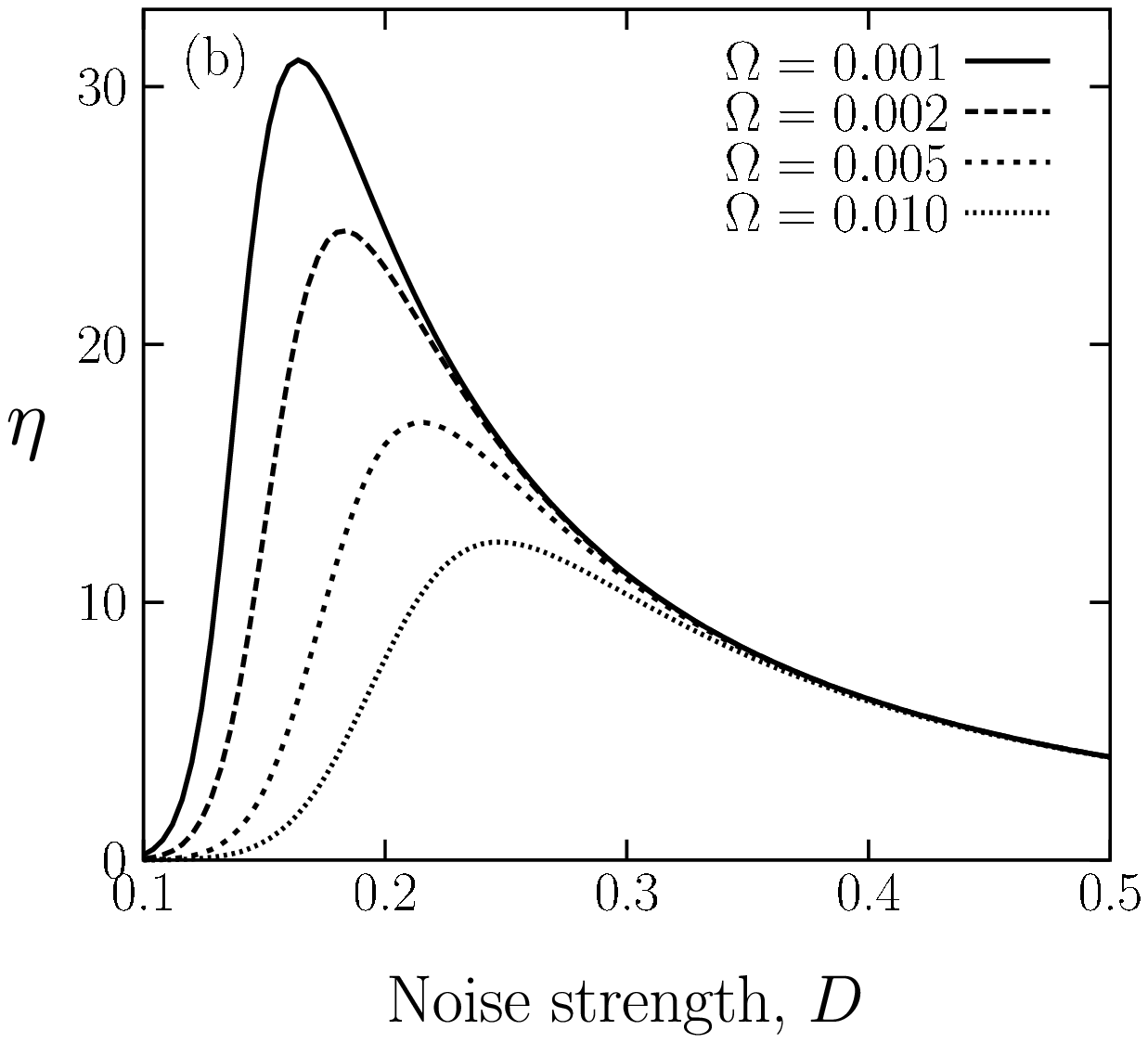,width=0.49\textwidth}

\caption{The spectral amplification of the signal (in arbitrary units)
is depicted versus the thermal noise intensity $D=k_BT$ at different driving frequencies
$\Omega$: (a)
non-Markovian symmetric system  and (b) its Markovian counterpart.
In the non-Markovian case, both RTDs
follow a Pareto law with $\gamma=0.2$. $D$ is scaled in units
of the barrier height $\Delta U$; $\Omega$ is scaled in units of $\nu_0$.}
\label{Fig2}
\end{figure}

In contrast to the overall simpler behavior of the spectral
amplification measure the SNR  displays prime new features,
 cf. Fig. 3 (a) and Fig. 3 (b). First, the SNR becomes
frequency-dependent. In the
limit $\Omega\to 0$, we obtain for the form-factor $N(\Omega)$ in
Eq. (\ref{N}),
\begin{eqnarray}\label{N1}
N(\Omega)\approx \frac{[\langle
\tau\rangle\Omega]^{1-\gamma}}{2\sin(\pi\gamma/2)
\gamma^{\gamma}\Gamma(1-\gamma)}.
\end{eqnarray}
In this limit, the signal-to-noise ratio can be
approximated as
\begin{eqnarray}\label{SNR4}
{\rm SNR}(\Omega,D)\approx \frac{\pi}{4}(f_0\Delta x/2)^2\frac{(2\nu_0)^{\gamma}}
{\sin(\pi\gamma/2)\gamma^{\gamma}\Gamma(1-\gamma)}\frac{\exp\left(
-\gamma\Delta U/D\right)}{D^2}\Omega^{1-\gamma}.
\end{eqnarray}
This SNR expression (\ref{SNR4}) displays several nontrivial
features: (i) the Stochastic Resonance peak occurs at the noise strength
$D_{NM}(\Omega\to 0) =\gamma
\Delta U/2$ as compare to the Markovian
case, where $D_{M}=\Delta U/2$. (ii) The SNR
displays a nontrivial, power law dependence on the angular driving
frequency
${\rm SNR}(\Omega)\sim \Omega^{1-\gamma}$. Moreover, with the increasing
angular frequency $\Omega$ of signal the signal-to-noise ratio,
${\rm SNR}(\Omega)$, should approach its frequency independent Markovian
limit. Thus, the resonance value
 $D_{NM}(\Omega)$ becomes frequency-dependent for
 an intermediate range of frequencies and approaches
the Markovian value $D_M$ in the limit of high frequencies.
This profound frequency-dependence of non-Markovian Stochastic
Resonance is very distinct from its Markovian
counterpart, compare Fig. 3(a) with Fig. 3(b).

\begin{figure}

\epsfig{file=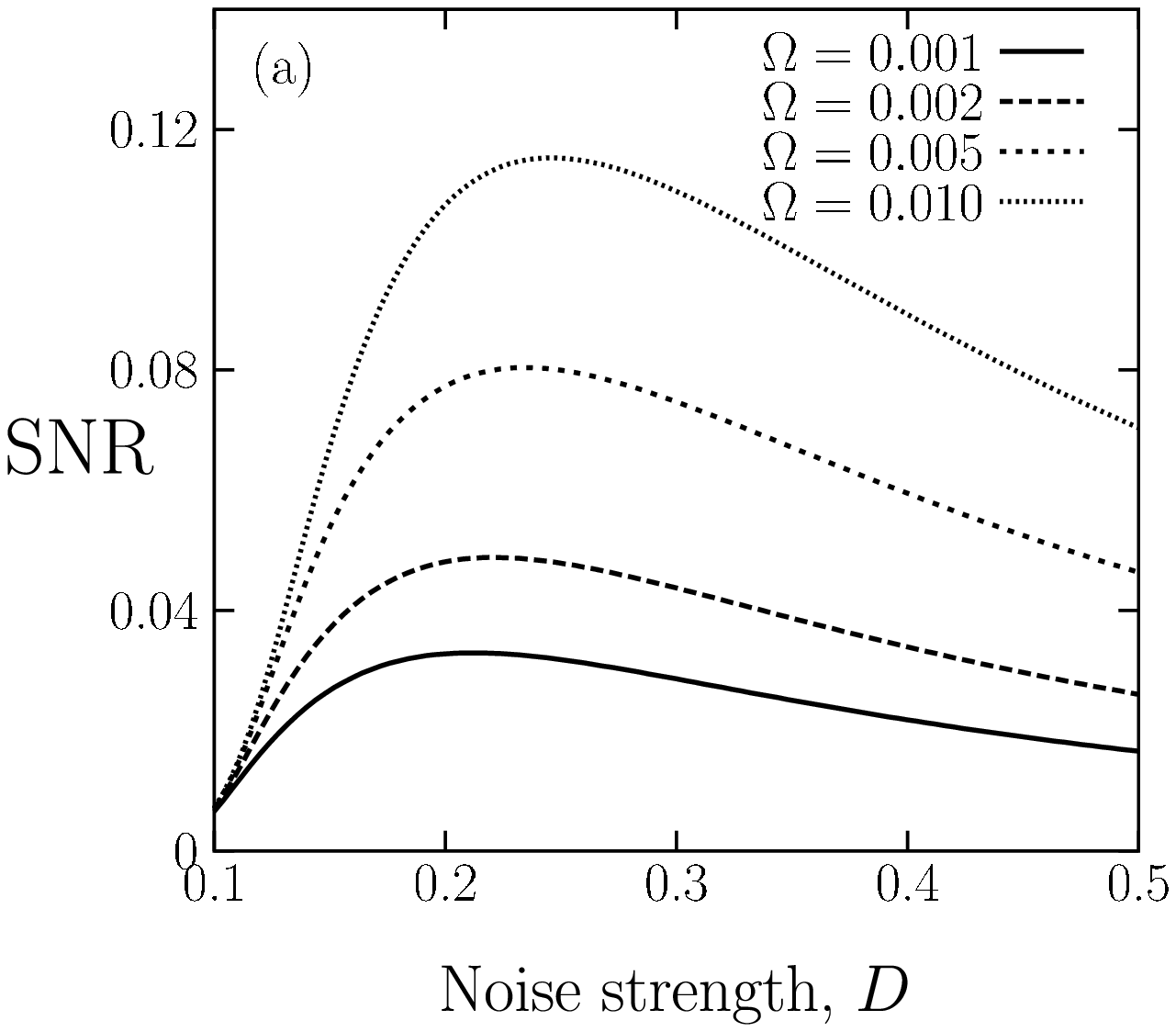,width=0.49\textwidth}
\hfill
\epsfig{file=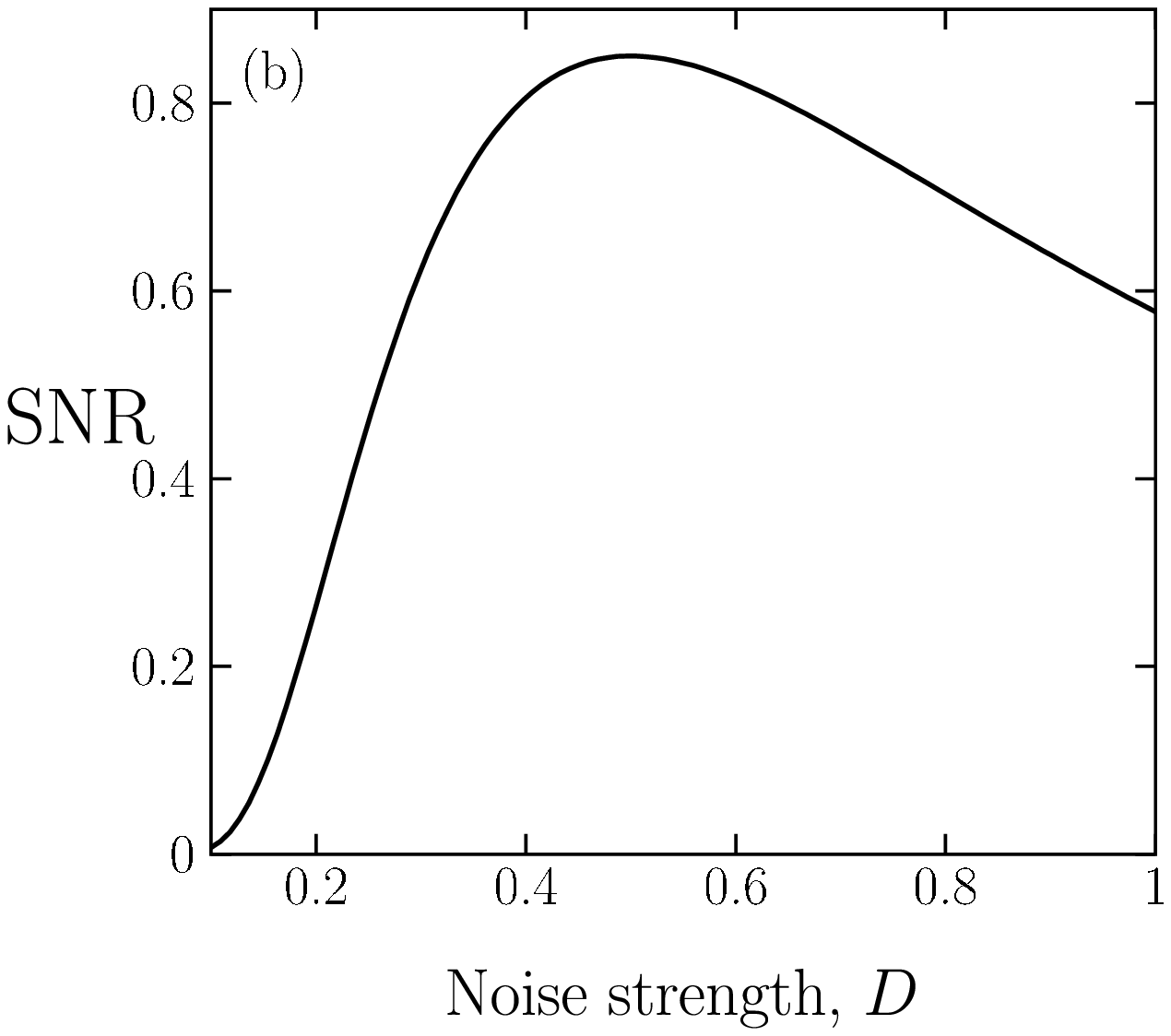,width=0.49\textwidth}

\caption{Signal-to-noise ratio (in arbitrary units)
versus thermal noise intensity $D=k_BT$ at different driving frequencies
$\Omega$: (a)
non-Markovian symmetric system  and (b) its Markovian counterpart.
In the non-Markovian case, both RTDs
follow a Pareto law with $\gamma=0.2$. $D$ is scaled in the units
of barrier height $\Delta U$, $\Omega$ is measured in units of $\nu_0$.}
\label{Fig3}
\end{figure}

\subsection{SR in ion channels with fractal kinetics}

Our second example pertains to the non-Markovian SR in an asymmetric system.
An especially interesting case emerges when one of the RTDs is exponential,
while the one presents a power law with a giant (divergent) dispersion.
Interestingly enough, such a case apparently is realized in nature for
the gating dynamics of the locust BK channel \cite{mercik}.
Indeed, this and some other ion channels
exhibit a fractal
gating kinetics together with the $1/f^{\alpha}$ noise power
spectrum of fluctuations
\cite{mercik,siwy,bezrukov2000,pnas}. In the context of gating dynamics,
$x(t)$ corresponds to the conductance fluctuations and the forcing $f(t)$
is proportional
to the time-varying transmembrane voltage. For a locust BK channel the
measured unperturbed
closed time statistics $\psi_1(\tau)$ can be approximated by
a Pareto law (\ref{Pareto})
with $\gamma\approx 0.24$ and $\langle \tau_1\rangle=0.84$ ms \cite{mercik}.
The open time RTD assumes an exponential form
with $\langle \tau_2\rangle=0.79$ ms \cite{mercik}.

Unfortunately, neither the voltage, nor the temperature
dependence of the mean residence times  are experimentally available. For
this reason,
we employ here the common Arrhenius
dependence in (\ref{Arrhenius}) with some characteristic parameters.
Namely, because the
temperature dependence of  open-to-closed transitions is typically
strong \cite{hille}, we assume a high activation barrier; i.e.
$\Delta U_2=100$ kJ/mol
($\sim 40\;k_BT_{room}$). The closed-to-open transitions are assumed to
be  weakly temperature-dependent with $\Delta U_1=10$ kJ/mol. Because
$\langle \tau_1 \rangle \sim \langle \tau_2 \rangle$ at room temperature
$T_{room}$,  the difference between
$\Delta U_1$ and $\Delta U_2$ is  compensated
by an entropy difference $\Delta S\sim
-36\;k_B$. The physical reasoning is that
the closed time statistics
exhibits a power law; i.e. the  conformations in the closed state
form a self-similar hierarchy and
are largely
degenerate \cite{pnas}.  This in turn implies
a larger entropy as compared to the open state.

The normalized autocorrelation function
$k(t)$ and the power spectrum $S_N(\omega)$ of the current
fluctuations are of prime interest. In the considered case, the auxiliary function
(\ref{aux}) simplifies to
\begin{eqnarray}\label{form}
\tilde G(s)=\frac{\langle \tau_2\rangle s[1-\tilde \psi_1(s)]}
{\langle \tau_2\rangle s+ 1-\tilde \psi_1(s)},
\end{eqnarray}
where $\tilde \psi_1(s)$ is given by Eq. (\ref{example1}) with
$\langle \tau\rangle =\langle \tau_1\rangle $.
The Laplace-transform of $k(t)$ can in the limit $s\to 0$
be approximated as
\begin{equation}\label{chan_ks}
 \tilde k(s) \to
\frac{\gamma^{\gamma}\Gamma(1-\gamma)\langle \tau_1\rangle
\langle \tau_2\rangle}{\langle \tau_1\rangle +
\langle \tau_2\rangle}\left[ \langle \tau_1\rangle s \right ]
^{\gamma-1}.
\end{equation}
From (\ref{chan_ks}) the long-time ($t\to\infty$) behavior of the
autocorrelation function follows immediately by virtue of a
Tauberian theorem \cite{hughes}, namely
 \begin{equation}\label{chan_t}
k(\tau)\to p_2^{st} \left(\frac{\tau}{\gamma \langle \tau_1\rangle} \right)
^{-\gamma},
\end{equation}
where $p_2^{st}$
is the channel's stationary opening probability. The result in
(\ref{chan_t})
describes a power law decay with an exponent $\gamma=0.24$.
In Fig. 4(a), this analytical result is compared with
the numerical
inversion of $\tilde k(s)$ with $\tilde G(s)$ in (\ref{form}),
obtained due to the Stehfest algorithm
\cite{stehfest}. This figure shows that the long-time
asymptotical behavior of $k(\tau)$ indeed obeys the power law in (\ref{chan_t})
for $\tau> 10\;{\rm sec}$. However, for smaller
$\tau<10\;{\rm sec}$
some kind of transient behavior occurs which cannot be characterized
by a simple power law. Nevertheless, the slow decay of correlations is
clearly non-exponential.

For $\omega\gg\langle \tau_{1,2}\rangle^{-1}$ the power spectrum of
fluctuations is expected to approach a Lorentzian tail,
$S(\omega)\sim \omega^{-2}$.
Indeed, this behavior starts in Fig. 4(b) for $\omega>500\; {\rm sec}^{-1}$.
The nontrivial frequency dependence
emerges for
the sufficiently small $\omega\ll \langle \tau_{1,2}\rangle^{-1}$.
In this case we obtain from (\ref{chan_ks})
\begin{eqnarray}\label{chan_power}
S(\omega\to 0)\approx 2 (\Delta x)^2 \frac{\langle \tau_1\rangle^2
\langle \tau_2\rangle^2}{(\langle \tau_1\rangle+
\langle \tau_2\rangle)^3}
\Gamma(1-\gamma)\sin(\pi\gamma/2)
\frac{\gamma}{[\gamma\langle \tau_1\rangle
\omega]^{1-\gamma}}.
\end{eqnarray}
Thus, for $\gamma=0.24$  we have $S(\omega\to 0)\propto
1/\omega^{\alpha}$ with $\alpha=1-\gamma=0.76$. This typical
$1/f^{\alpha}$ noise behavior is depicted in Fig. 4(b).
We should remark, however, that the experiment \cite{siwy} gives a slightly
different value of $\alpha\approx 1$. The reasons of this discrepancy
are presently not clear. One  possibility is that the durations of
the subsequent open and closed time intervals are yet mutually
correlated, contrary to the assumptions made in the present model.
If this is the case indeed, the studied model
should be generalized further to account for
such correlations.

\begin{figure}

\epsfig{file=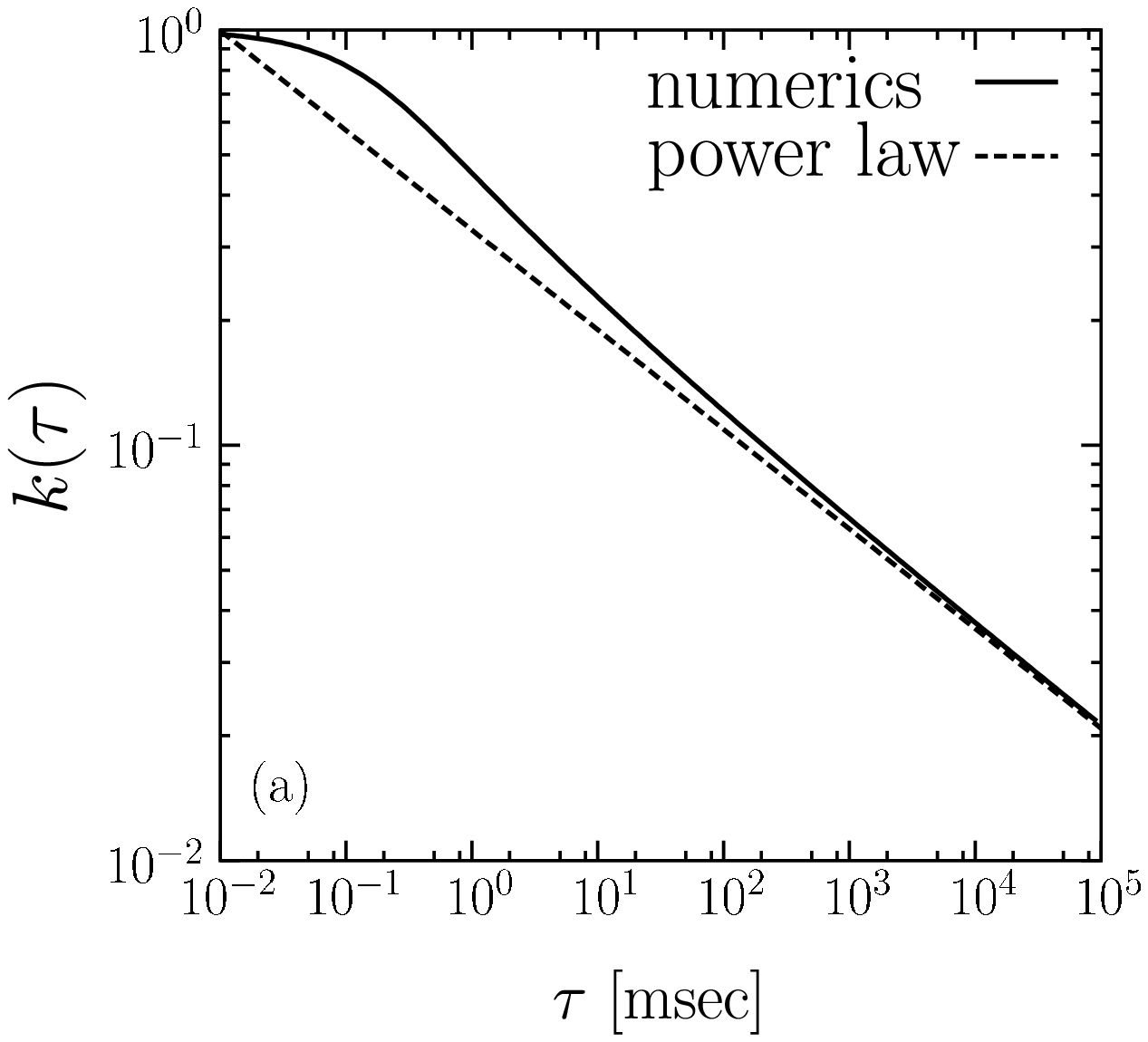,width=0.50\textwidth}
\hfill
\epsfig{file=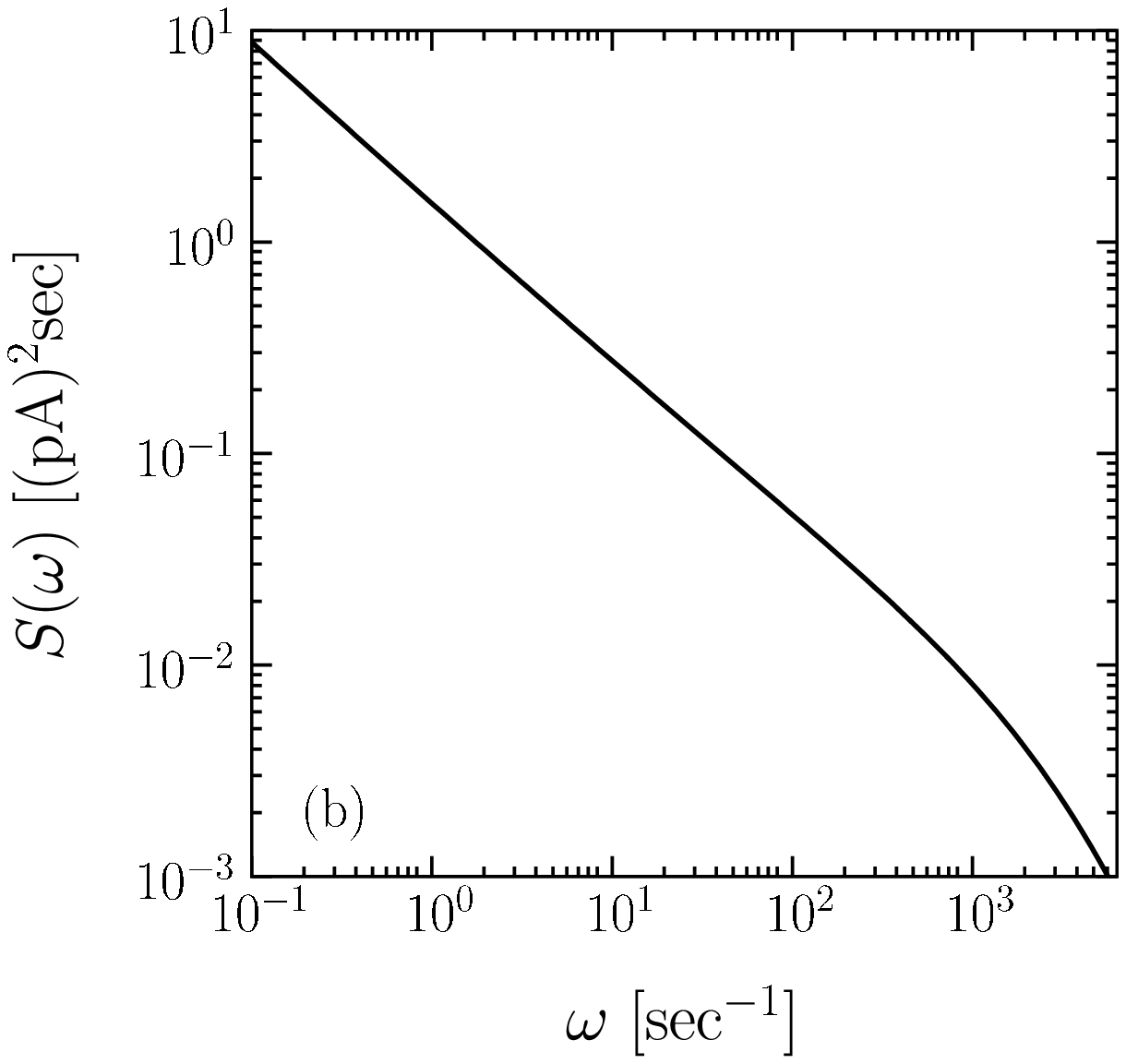,width=0.48\textwidth}

\caption{(a) The normalized autocorrelation function of current
fluctuations, see Eq. (\ref{correlation}), and
(b) the
corresponding power spectrum for the studied model of
locust BK channel. The amplitude of current fluctuations
is taken to be $10$ pA. The broken line in (a) corresponds to the long-time
asymptotic, Eq. (\ref{chan_t}), being in agreement
with the numerical result (full line) in long-time limit.}
\label{Fig4}
\end{figure}

\begin{figure}
\begin{center}
\epsfig{file=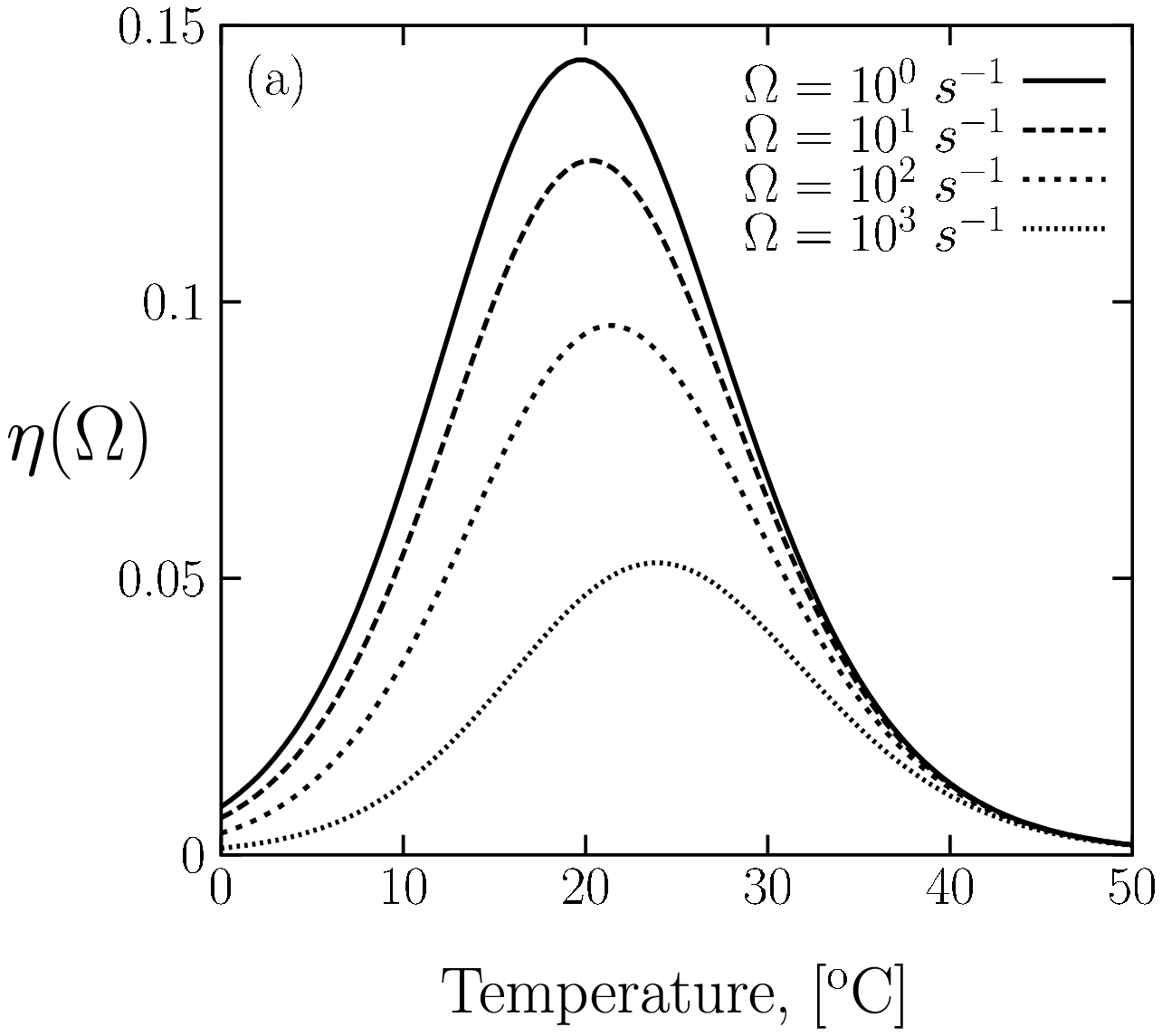,width=0.49\textwidth}
\hfill
\epsfig{file=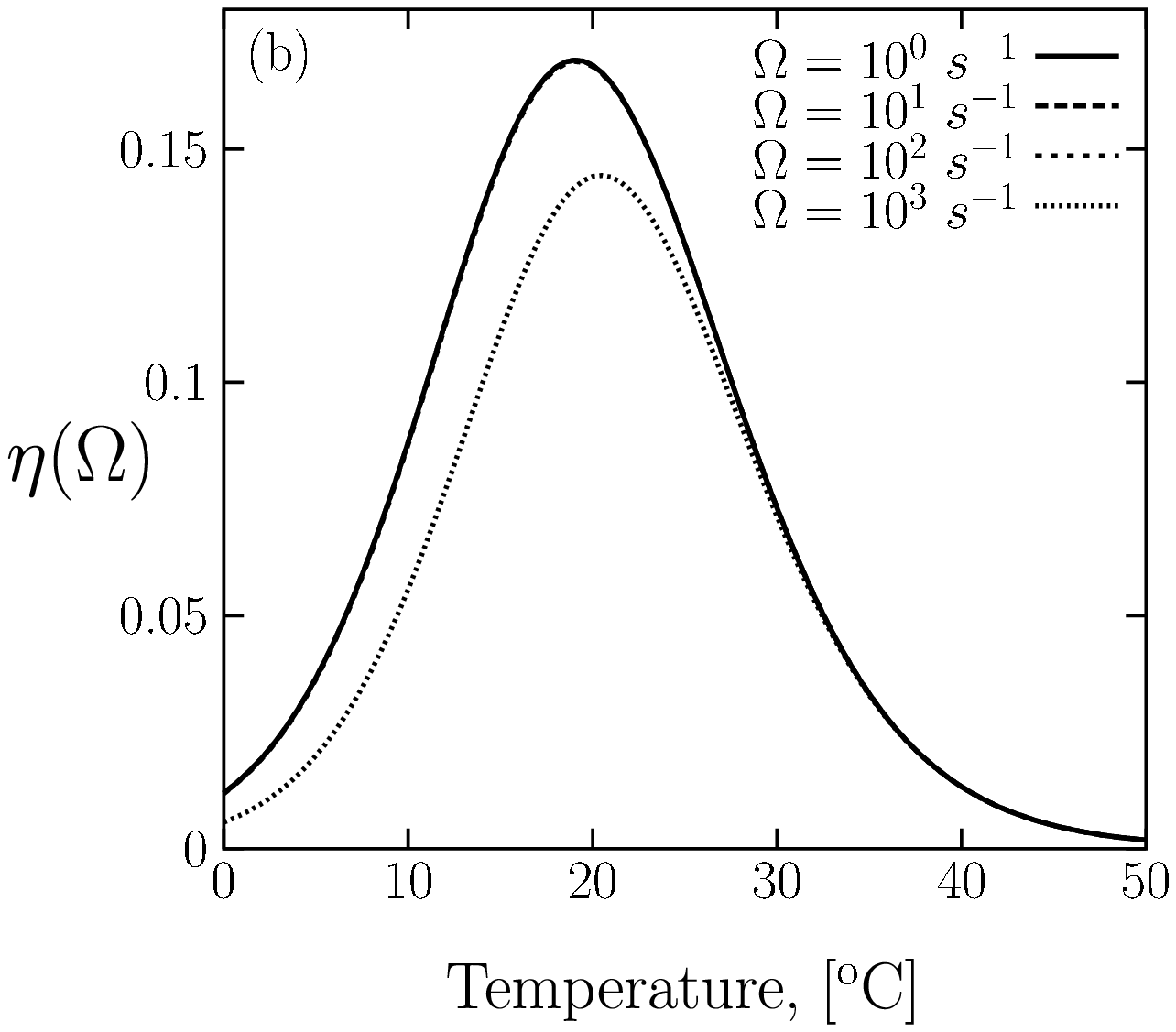,width=0.49\textwidth}
\end{center}
\caption{(a) The spectral power amplification  $\eta(\Omega)$, Eq. (\ref{res1}),
(in arbitrary units) {\it vs.} temperature (in $^o$C) for
the BK ion channel gating scenario (see text) and (b) its comparison with a corresponding Markov
modeling. }
\label{Fig5}
\end{figure}

The spectral power amplification  versus the
temperature
is depicted  for various angular
driving frequencies in Fig. 5(a). The panel in
Fig. 5(b)
corresponds to an overall  Markovian modeling with an exponential
$\psi_1(\tau)$
possessing the same mean residence time $\langle \tau_1 \rangle$.
We observe a series of
striking non-Markovian features
in Fig. 5: (i) A characteristic SR-maximum occurs in
the physiological range
of varying temperatures. This maximum is caused by entropic effects
which have been not addressed before in the theory of stochastic
resonance. Because of the fact that the free-energy bias $\epsilon(T)$
is temperature-dependent, due to a large entropic asymmetry between
states, Stochastic Resonance in the spectral amplification occurs
in a temperature regime where
the populations of both states become approximately equal,
$\epsilon(T)\approx 0$. Note that this effect occurs
also in the Markovian case, cf. Fig. 5(a) and Fig. 5(b). Therefore,
it is not caused by non-Markovian effects.
(ii) Due to
 a  intrinsic  asymmetry the (angular)
 frequency dependence of the spectral amplification
$\eta(\Omega,T)$ for the Markov modeling is rather weak for small
frequencies $\Omega\ll \langle \tau_{1,2}\rangle^{-1}$ \cite{review1}. In
contrast, the non-Markovian SR exhibits a distinct low-frequency
dependence, thereby frequency-resolving the three overlapping lines
 in Fig. 5b. This feature constitutes an authentic non-Markovian effect.
 (iii) The evaluation of the SNR yields --
 in clear contrast to the frequency-independent Markov
 modelling -- a profound, very strong non-Markovian 
 SR frequency {\em suppression} of SNR
 towards smaller frequencies: The SNR-maximum
 for the top line in Fig. 5(a) is suppressed by two orders of magnitude
  as compared to the Markov case,  cf. Fig. 6. As a consequence, for a
  strong non-Markovian situation
  it is preferable to use low-to-moderate frequency inputs in order
  to monitor non-Markovian Stochastic Resonance with SNR.

\begin{figure}

\epsfig{file=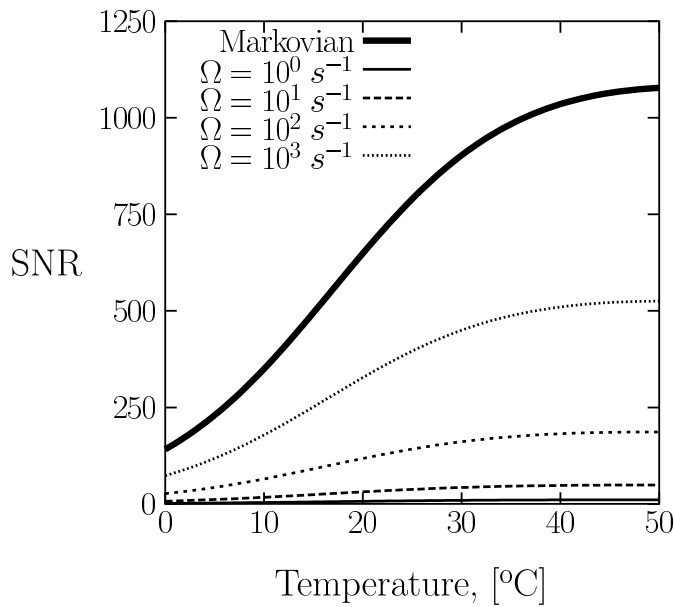,width=0.49\textwidth}

\caption{The signal-to-noise ratio (SNR in arbitrary units) versus
temperature (in $^o$C) for the studied
model of Stochastic Resonance in a locust BK channel. The upper curve depicts
the Markovian limit attained for large angular driving
frequencies of the signal.}
\label{Fig6}
\end{figure}

\section{Summary and conclusions}

In the present work we have put forward a general theory of
Stochastic Resonance for two state non-Markovian  systems.
The theory is based on time-inhomogeneous integral renewal equations
governing the evolution of conditional probabilities
in the presence of driving signal. These equations for driven
renewal processes  generalize earlier
result by Cox \cite{cox} and others \cite{boguna}
for stationary renewal processes to
include the signal influence on the residence time
distributions. Based on these new equations
we presented a general outline of the theory of the linear
and the asymptotic nonlinear response to the sinusoidal signal.
In particular, we obtained a general expression
for the linear response function $\tilde \chi(\omega)$,
Eq. (\ref{mainres}),
which can be used for a variety of applications.
The expression in
(\ref{mainres}) presents major result
of this paper.
We note, however, that the explicit use
of Eq. (\ref{mainres}) requires one to
specify explicitly the way in which the periodic signal modulates
the asymptotic, non-equilibrium residence time distributions.
For a class of non-equilibrium fractal
distributions where the signal enters  the RDTs through a single
frequency parameter having the meaning of the inverse mean residence
time, it has been shown that
Eq. (\ref{mainres}) reduces to the result (\ref{FTfreq2})
of the phenomenological
theory of linear response developed previously
in Ref. \cite{goychuk03}.
Moreover, if the mean residence times obey the thermal
detailed balance condition (\ref{detbal}), the expression (\ref{FTfreq2})
reduces further to Eq. (\ref{FTfreq3}) which can be obtained
independently
from the classical fluctuation-dissipation theorem (\ref{equilibrium})
by use of the expression in (\ref{laplace-corr}) for the autocorrelation
function of the considered non-Markovian stochastic process. Even though the
microscopic (or mesoscopic) details of the thermal equilibrium dynamics leading
to the observed two state non-Markovian fluctuations are generally
not known, the linear response function is determined uniquely by the
characteristic functions of the residence time-distributions
$ \tilde \psi_{1,2}(s)$ via Eqs. (\ref{FTfreq3}) and (\ref{aux}).
For such equilibrium non-Markovian fluctuations, the knowledge
of the equilibrium RTDs allows one to determine the linear response
of the considered physical system to weak signals.
This is the essence of the phenomenological theory
of non-Markovian stochastic resonance put forward in Ref. \cite{goychuk03}.
For such equilibrium systems, the general expressions for
the spectral power amplification, Eq. (\ref{res1}), and for the
signal-to-noise ratio (SNR), (\ref{res3}), are available.
 We applied these general expressions to study the main features
of Stochastic Resonance in several non-Markovian systems
exhibiting long-range temporal correlations
along with $1/f^{\alpha}$ power spectra of fluctuations.

In particular, for a symmetric non-Markovian system with a
power law distributed residence time intervals
the occurrence of Stochastic Resonance
has been demonstrated to comply with a stochastic frequency
synchronization similar to the Markovian case \cite{review1}.
However,  both the SPA-measure and the SNR-measure become strongly
suppressed due to strong non-Markovian effects. The most striking feature
of the non-Markovian SR is a distinct {\it frequency dependence
of the SNR measure}. In particular, the SNR becomes immensely
suppressed for low frequency signals. Thus, the
use of signals with an intermediate frequency range matching the
mean time  of the stochastic escapes between states yields
most distinct non-Markovian SR-feature.

For asymmetric non-Markovian fluctuations pertinent
to fractal gating dynamics of the locust BK ion channel
several new features have been revealed. (i) The expected
diminishment of the SPA-measure relative to the Markovian case
does not occur. This can be attributed to the fact that one of the
RDTs in the considered case is strictly exponential
similar the Markovian case. (ii) For asymmetric
Markovian systems the SPA-measure ceases to be frequency-dependent
for small adiabatic frequencies. The non-Markovian effects, however,
introduce at low driving frequencies a distinct dependence,
both for the SPA and the SNR. This latter phenomenon can be used
to detect and establish a strong non-Markovian behavior in practice.

Our novel non-Markovian theory of Stochastic Resonance possesses a whole
range of applications
and we hope that it will be used by the practitioners
in their further research work on Stochastic Resonance.
Especially, we hope that our
theory will guide experimentalists to find the proper
and most interesting
parameter regimes and to reveal the Stochastic Resonance
effect on the level of single biomolecules.

\acknowledgements

This work has been supported by the Deutsche Forschungsgemeinschaft
within SFB 486
``Manipulation of matter on the nanoscale'', project A10.


\begin{references}
\bibitem{remark0} This observation can be rationalized as follows.
Let us consider a sojourn in one state (say state $x_1$)
characterized by the survival
probability
$\Phi_1(\tau)=\int_{\tau}^{\infty}\psi_1(t)dt$. The
corresponding
residence time interval $[0,\tau]$ can be arbitrarily
divided into two pieces
$[0,\tau_1]$
and $[\tau_1,\tau]$.
If no memory effects are present, then
$\Phi_1(\tau)=\Phi_1(\tau-\tau_1)\Phi_1(\tau_1)$.
The only nontrivial solution of this latter functional equation which decays
in time reads $\Phi_1(\tau)=\exp(-\nu_1\tau)$, with $\nu_1>0$.
This corresponds to a Markovian case.
Otherwise, the process $x(t)$ is not Markovian and the master equation
(\ref{rate-eq}) must be substituted by a non-Markovian master equation.
\bibitem{remark6} This statement is valid only for continuous RTDs. It
does not apply when RTD possesses a
discontinuity, e.g., for $\psi(\tau)=\delta(\tau-\tau_0)$. Such
special cases are not considered here.
\bibitem{remark7} One should stress however that in any {\it
realistic} situation
$S_N(\omega=0)$ should be finite, i.e. there should always be a low frequency
cutoff $\omega_{low}$ present such that $S_N(\omega)\approx S_N(0)=const$ for
$\omega<\omega_{low}$. The experimental time-series having
$1/\omega^{\alpha}$ spectrum can be considered as strictly
stationary only if
the observation time exceeds $1/\omega_{low}$.
\end{references}
\end{document}